\newcommand{\NPA}[3]{Nucl.\ Phys.\ A\ {\bf #1},\ #2 (#3)}
\newcommand{\NPB}[3]{Nucl.\ Phys.\ B\ {\bf #1},\ #2 (#3)}

\newcommand{\PLB}[3]{Phys.\ Lett.\ B\ {\bf #1},\ #2 (#3)}

\newcommand{\PRL}[3]{Phys.\ Rev.\ Lett.\ {\bf #1},\ #2 (#3)}

\newcommand{\PRC}[3]{Phys.\ Rev.\ C\ {\bf #1},\ #2 (#3)}
\newcommand{\PRD}[3]{Phys.\ Rev.\ D\ {\bf #1},\ #2 (#3)}
\newcommand{\JPG}[3]{J.\ Phys.\ G\ {\bf #1},\ #2 (#3)}

\newcommand{\ZPC}[3]{Z.\ Phys.\ C\ {\bf #1},\ #2 (#3)}

\newcommand{\PTP}[3]{Prog.\ Theo.\ Phys.\ {\bf #1},\ #2 (#3)}

\newcommand\f{\phi}

\newcommand{\diracslash}[1]{#1\llap{/\kern2pt}}

\newcommand{\be}{\begin{equation}}
\newcommand{\ee}{\end{equation}}
\newcommand{\bea}{\begin{eqnarray}}
\newcommand{\eea}{\end{eqnarray}}
\newcommand{\ba}[1]{\begin{array}{#1}}
\newcommand{\ea}{\end{array}}

\documentclass[prd,aps,floats,nofootinbib,tightenlines,showpacs]{revtex4-1}
\usepackage{epsfig,graphicx,pstricks}
\usepackage{psfrag}
\usepackage{color}
\usepackage{amsmath}
\usepackage{amsfonts}
\usepackage{amssymb}
\usepackage{textcomp}
\usepackage{multirow}

\begin{document}

\title {CP violation  and chiral symmetry breaking in hot and dense quark matter
in presence of magnetic field}
\author{Bhaswar Chatterjee}
\email{bhaswar@vecc.gov.in}
\affiliation{Department of Physics, Variable Energy Cyclotron Center, Kolkata 700064, India}
\author{Hiranmaya Mishra}
\email{hm@prl.res.in}
\affiliation{Theory Division, Physical Research Laboratory,
Navrangpura, Ahmedabad 380 009, India}
\author{Amruta Mishra}
\email{amruta@physics.iitd.ac.in}
\affiliation{Department of Physics, Indian Institute of Technology, New 
Delhi 110016, India}

\date{\today} 

\def\be{\begin{equation}}
\def\ee{\end{equation}}
\def\bearr{\begin{eqnarray}}
\def\eearr{\end{eqnarray}}
\def\zbf#1{{\bf {#1}}}
\def\bfm#1{\mbox{\boldmath $#1$}}
\def\hf{\frac{1}{2}}
\def\sl{\hspace{-0.15cm}/}
\def\omit#1{_{\!\rlap{$\scriptscriptstyle \backslash$}
{\scriptscriptstyle #1}}}
\def\vec#1{\mathchoice
        {\mbox{\boldmath $#1$}}
        {\mbox{\boldmath $#1$}}
        {\mbox{\boldmath $\scriptstyle #1$}}
        {\mbox{\boldmath $\scriptscriptstyle #1$}}
}

\begin{abstract}
We investigate chiral symmetry breaking and strong CP violation effects on the phase diagram of 
strongly interacting matter in presence of a constant magnetic field. The effect of magnetic field
and strong CP violating term on the phase  structure at finite temperature and density is studied
within a three flavor Nambu-Jona-Lasinio (NJL) model  
including the Kobayashi-Maskawa-t'Hooft (KMT) determinant term.
This is investigated using an explicit variational ansatz for ground state with quark anti-quark
pairs leading to condensates both in scalar and pseudoscalar channels. Magnetic field enhances the 
condensate in both the channels. Inverse magnetic catalysis for CP transition at finite chemical potential is
seen for zero temperature and for small magnetic fields.
\end{abstract}

\pacs{12.38.Mh, 12.39.-x, 11.30.Rd, 11.30.Er}

\maketitle

\section{Introduction}
The study of Charge-Parity ($CP$) violation in strong interaction is of immense importance in the 
context of early universe scenario\cite{pecceiquinn,cpcosmo} 
as well as heavy ion
 collision experiments \cite{dimacp,cme,starexp}.
 Strong interaction respects 
space time reflection symmetry
to a very high degree. However, this is not a direct consequence of quantum chromodynamics (QCD). The existence of
instanton configurations for QCD allows for a nontrivial
topological term in the action, the so-called $\theta$-term of QCD Lagrangian given as

\begin{equation}
{\cal L}_\theta=\frac{\theta}{64\pi^2} g^2 F_{\mu\nu}^a\tilde F^{a\mu\nu}.
\label{lth}
\end{equation}

\noindent In the above, $F_{\mu\nu}^a$ is the gluon field strength and $\tilde F^{\mu\nu}$ being its dual. This term violates charge conjugation and parity unless $\theta=0$ and $\pm \pi$.  However, experiments on neutron dipole moment set limit
on the value of $\theta$
as 
 $\theta < 0.7\times 10^{-11}$ \cite{endm}. This smallness
of the CP violation term or its complete absence is 
not understood completely though a possible
explanation is given in terms of spontaneous breaking of a new symmetry
 the Peccei-Quinn symmetry\cite{pecceiquinn}. In  vacuum,  i.e.
for zero temperature and zero density, parity is preserved when $\theta=0$
exactly \cite{vafawit}. However, this is spontaneously broken at $\theta=\pi$\cite{dashen}. This P violation,
 called the Dashen mechanism is essentially nonperturbative.

Even if  CP 
is not violated for QCD vacuum, it is conceivable  that it can be violated 
for QCD matter at
finite temperature or density. In deed, it has been proposed that 
hot matter produced in heavy ion collision
experiments can give rise to domains of meta stable states that violate CP
  locally \cite{dimacp}. Different 
experimental observables for detecting
such a phase have been suggested \cite{dimapisarski}. Apart from producing
high temperature, colliding nuclei also produce transient
strong magnetic fields. A nonzero $\theta$ leads to a deviation of left and right handed helicity quarks.
 As a consequence an electromagnetic current is generated along the magnetic field. Such a mechanism  known
 as chiral magnetic effect
(CME)\cite{cme}
 may explain the charge
separation in the recent STAR results \cite{starexp}.
This makes the study of chiral symmetry breaking mechanism at finite $\theta$
interesting at finite temperature and magnetic fields.
On the other hand, in the context of cold and dense matter, compact stars can be strongly 
magnetized. The magnetars, which are strongly magnetized neutron stars
may have strong magnetic fields of the order of $10^{15}$--$10^{16}$ gauss 
 \cite{dunc,duncc,dunccc,duncccc,lat,broder,lai} . 

In the present work we intend to investigate how chiral transition is affected
when the CP odd effects  and a strong magnetic background is present for hot and dense matter. For this purpose,
 we adopt the three flavor Nambu Jona-Lasinio model 
as an effective theory for chiral transitions. The
effect of axial anomaly and the strong CP violation here is included
through  the Kobayashi-Masakawa-t'Hooft determinant term that mimics
the effects of nontrivial gauge field configuration. This term is also
a function of $\theta$ and is responsible for CP violation for
non vanishing values of $\theta$. Such a term has been extensively studied
earlier for NJL models with two flavors in Ref.s \cite{cpnjl,bbone,bbtwo} for
 studying the effects of non zero $\theta$ on the chiral transition.
This has been further extended to the two flavor NJL model including
Polyakov loop potential\cite{sakai,kuonop}. We had earlier considered
the effects of strong CP violation on chiral symmetry breaking
for the realistic case of 2+1 flavor using the NJL model \cite{bhaswarcp}.
This was further extended in Ref.\cite{kuonop} to include the effects of  Polyakov loop
potential. In all these investigations the effects of magnetic field
were not included. It is this question that we would like to investigate here.
 
Modification of the ground state of QCD for $\theta=0$ in connection with chiral symmetry
breaking in presence of magnetic field has been investigated in different effective
models- e.g. chiral perturbation theory\cite{chiralB}, NJL model \cite{NJLB,bhaswarnjlb,providencia,boomsma}
as well as different quark models of hadrons.
In various models it was seen that while magnetic field acts as a catalyser of chiral symmetry breaking,
it was observed that medium effects can lead to inverse magnetic catalysis for the same particularly at
finite chemical potentials. The effects of magnetic field as well as nonzero values for
$\theta$ has been considered in Ref.\cite{fragalsm} within the chiral sigma model.

We organize the paper as follows. In section II, we consider the three flavor NJL model along with the
 CP violating theta dependent six fermion determinant interaction term that also  breaks axial symmetry .
Here, we also write down quark field operator expansions in presence of magnetic field.
Using the same, we next consider a variational ground state with quark anti-quark pairs that
is related to chiral symmetry breaking. The ansatz is taken general enough to include both scalar as well as
pseudoscalar condensates. The pseudoscalar condensate takes nonzero values for
finite values of $\theta$. In section III, we discuss the resulting phase diagram at finite
temperature as well as finite density for different strengths of magnetic field and for various values of $\theta$.
Finally, in section IV we summarize the results and conclusion with a possible outlook.

\section{NJL model with CP violation and an ansatz for the ground state}

To describe the chiral phase structure of strong interactions including the CP violating effects, 
we use the 3-flavor NJL model along with the flavor mixing determinant term. The Lagrangian is 
given by

\begin{equation}
{\cal L} = \bar\psi\left(iD\sl - m\right)\psi
+ G\sum_{A=0}^8\left[(\bar\psi\lambda^A\psi)^2 + (\bar\psi i\gamma^5\lambda^A\psi)^2\right]
- K\left[e^{i\theta}det\lbrace\bar\psi(1+\gamma^5)\psi\rbrace +
e^{-i\theta}det\lbrace\bar\psi(1-\gamma^5)\psi\rbrace\right],
\label{lag3fl}
\end{equation}

\noindent
where $\psi ^{i,a}$ denotes a quark field with color `$a$' 
$(a=r,g,b)$, and flavor `$i$'
 $(i=u,d,s)$, indices.  $D_\mu=\partial_\mu-iqA_\mu$ is the 
covariant derivative in the presence of external magnetic field $\zbf B$ 
which we assume to be constant and in the $z$- direction. Further,
we choose the gauge such that the corresponding electromagnetic potential
is given as $A_\mu=(0,0,Bx,0)$.
The matrix of current quark masses is given by
$\hat m$=diag$_f(m_u,m_d,m_s)$ in the flavor space.
We  shall assume  in the present investigation, isospin
symmetry with $m_u$=$m_d$.  
Strictly speaking, when the electromagnetic effects are taken into account,
the current quark masses of u and d quarks should not be the same due to the
difference in their electrical charges. However, because of the smallness of the
electromagnetic coupling, we shall ignore this tiny effect and continue with
$m_u=m_d$ in the present investigation of chiral symmetry breaking in strong interaction.
In Eq. (\ref{lag3fl}), $\lambda^A$, $A=1,\cdots 8$ denotes the Gellman matrices
acting in the flavor space and
$\lambda^0 = \sqrt{\frac{2}{3}}\,1\hspace{-1.5mm}1_f$,
$1\hspace{-1.5mm}1_f$ as the unit matrix in the flavor space.
The four point interaction term $\sim G$ is symmetric in $SU(3)_V\times
SU(3)_A\times U(1)_V\times U(1)_A$. In contrast, 
the determinant term $\sim K$, which generates
a six point interaction for the case of three flavors, breaks $U(1)_A$ symmetry for vanishing
$\theta$ values.
 In the absence of the magnetic field and
the mass term, the overall symmetry in the flavor space 
is $SU(3)_V\times SU(3)_A \times U(1)_V$. This
spontaneously breaks to $SU(3)_V \times U(1)_V$ implying
the conservation of the baryon number and the flavor number. The current
quark mass term introduces additional explicit breaking of chiral symmetry
leading to partial conservation of  the axial  current. Due to the
presence of magnetic field, on the other hand,  the $SU(3)_V$ symmetry
 in the flavor space reduces to to $SU(2)_V\times SU(2)_A$ since the u quark
 has different  electric charge compared
 to d and s quarks \cite{manfer}.
 The effect of CP violating topological term of Eq.(\ref{lth}) is
 simulated by the determinant
term of Eq.(\ref{lag3fl}) in the quark sector.
This can be easily seen by
taking the divergence of the flavor singlet axial current
\be
\partial_\mu J_5^\mu=2i\bar\psi m\gamma^5\psi+2i N_fK
\left(e^{i\theta}det\bar\psi(1+\gamma^5)\psi -h.c.\right),
\ee
where, $J_5^\mu=\bar\psi\gamma^\mu\gamma^5\psi$ summed over all the flavors. This equation
may be compared with the usual anomaly equation written in terms of the topological
term for the gluon field arising from Eq.(\ref{lth})
\be
\partial_\mu J_5^\mu=2i\bar\psi m\gamma^5\psi
+2 N_f\frac{\theta}{32\pi^2} g^2 F_{\mu\nu}^a\tilde F^{a\mu\nu}.
\ee
Thus the effect of gluon operator $\frac{\theta}{32\pi^2} g^2
 F_{\mu\nu}^a\tilde F^{a\mu\nu}$ is simulated by the imaginary part of the
determinant term in the quark sector. Such a term can lead to formation of 
condensates in the pseudoscalar channel as we investigate in the following.

 The quark field operators in presence of a constant magnetic field can be 
expanded in terms of creation and annihilation operators can be written as
 as\cite{bhaswarnjlb}
\be
 \psi(\zbf x) = \sum_n\sum_r\frac{1}{2\pi}\int{d\zbf p_{\omit x}\left[q_r(n,\zbf p_{\omit x})
U_r(x,\zbf p_{\omit x},n) + \tilde q_r(n,-\zbf p_{\omit x})V_r(x,-\zbf p_{\omit x},n)\right]
e^{i\zbf p_{\omit x}\cdot\zbf x_{\omit x}}}.
\label{psiex}
\ee
The sum over the integers $n$ in the above expansion runs from 0 to infinity.
In the above,
$\zbf p_{\omit x}\equiv (p_y,p_z)$, and, $r=\pm 1$ denotes
the up and down spins. The two component quark and anti quark operators satisfy 
the quantum algebra

\begin{equation}
 \lbrace q_r(n,\zbf p_{\omit x}),q_{r^\prime}^\dag(n^\prime,\zbf p_{\omit x}^\prime)
\rbrace =
\lbrace \tilde q_r(n,\zbf p_{\omit x}),\tilde q_{r^\prime}^\dag(n^\prime,\zbf p_{\omit x}^\prime)\rbrace =
\delta_{rr^\prime}\delta_{nn^\prime}\delta(\zbf p_{\omit x}-\zbf p_{\omit x}^\prime).
\label{acom}
\end{equation}
Further, $U$ and $V$ are the four component spinors for the quark and anti quarks respectively. 
For constant magnetic field they have been derived in Ref.\cite{bhaswarnjlb}. These can be expressed in terms of the
 Hermite polynomials and are normalized as \cite{bhaswarnjlb}
\be
\int dx U_r(x,\zbf p_{\omit x},n)^\dagger U_s(x,\zbf p_{\omit x},m)
=\delta_{n,m}\delta_{r,s}
=\int dx V_r(x,\zbf p_{\omit x},n)^\dagger V_s(x,\zbf p_{\omit x},m)
\label{normspinor}
\ee

With the operators defined for the quark fields in a constant magnetic field, we next consider an ansatz
for the ground state as
\be
|\Omega\rangle=U|0\rangle
\label{ansatz}
\ee

where, $U_q=U_{I}U_{II}$ is an unitary operator. $U_{I}$ and $U_{II}$ are unitary operators
described in terms of quark-anti quark creation and annihilation operators. Explicitly they are
given as

\begin{equation}
U_{I}=
\exp\left(
\sum_{n=0}^{\infty} \int{d\vec p_{\omit x} 
{q_r^i}^\dag(n,\vec p_{\omit x})a_{r,s}^i(n,p_z)f^i
(n,\vec p_{\omit x})
\tilde q_s^i(n,-\vec p_{\omit x})} -h.c.\right)
\label{ansatz1}
\end{equation}
where, we have retained flavor index $i$ for the quark field operators.
Further, in the above equation, the spin dependent structure  $a_{r,s}^i$ is given by \cite{bhaswarnjlb}
\begin{equation}
 a_{r,s}^i=\frac{1}{|\zbf p^i|}\left[-\sqrt{2n|q^i|B}\delta_{r,s}-ip_z\delta_{r,-s}\right]
\end{equation}
with $|\zbf p^i| = \sqrt{p_z^2+2n|q^i|B} $ denoting the magnitude
of the three momentum of the quark/anti quark of $i$-th flavor
(with electric charge $q^i$) in presence of a magnetic field.
It is easy to show
that, $a a^\dagger=I$, where
$I$ is an identity matrix in two dimensions. The ansatz functions $f^i(n,p_z)$
are determined from the minimization of thermodynamic potential. This particular
ansatz of Eq.(\ref{ansatz1}) is a direct generalization of the ansatz
considered earlier \cite{hmspmnjl},  for vacuum structure for
chiral symmetry breaking to include the effects of magnetic field.
Next, the unitary operator $U_{II}$  is  given as
\begin{equation}
U_{II}=
\exp\left(
\sum_{n=0}^{\infty} \int{d\vec p_{\omit x} 
{q_r^i}^\dag(n,\vec p_{\omit x})r g^i
(n,\vec p_{\omit x})
\tilde q_s^i(n,-\vec p_{\omit x})} -h.c.\right)
\label{ansatz2}
\end{equation}
The above construct of Eq.(\ref{ansatz}) is a generalization of the ground state structure in presence of a CP violating term 
same in Ref\cite{bhaswarcp} to include  the effects of a nonvanishing constant  magnetic field as well.
Clearly, in Eq.(\ref{ansatz}) the  ansatz for the ground state has two arbitrary functions $f^i$ and $g^i$
which will be related to the condensates in the scalar and pseudoscalar channel respectively.  The effect of temperature
and density can also be implemented with such a nontrivial structure for the ground state using the formalism 
of thermo field dynamics
\cite{tfd,amph4}. Here, the statistical average of an operator is given as an expectation value
over a `thermal vacuum'. The methodology of TFD involves the doubling of the Hilbert space
\cite{tfd}. Explicitly, the `thermal vacuum' is constructed from the ground state at zero
temperature and density through a thermal Bogoliubov transformation given as

\begin{equation}
|\Omega(\beta,\mu)\rangle = {\cal U}_F|\Omega\rangle = 
e^{{\cal B}(\beta,\mu)^\dagger-{\cal B}(\beta,\mu)}|\Omega\rangle
\label{ubt}
\end{equation}

with,
\begin{equation}
{\cal B}^\dagger(\beta,\mu) =\int \Big [
\sum_{n=0}^{\infty} \int d\vec k_{\omit x}
q_r^\prime (n,k_z)^\dagger \theta_-(k_z,n, \beta,\mu)
\underline q_r^{\prime} (n,k_z)^\dagger +
\tilde q_r^\prime (n,k_z) \theta_+(k_z,n,\beta,\mu)
\underline { \tilde q}_r^{\prime} (n,k_z)\Big ].
\label{bth}
\end{equation}
In Eq.(\ref{bth}),
the underlined operators are the operators in the extended Hilbert space
 associated with thermal doubling in TFD method, and,
 the ansatz functions $\theta_{\pm}(n,k_z,\beta,\mu)$
are related to quark and anti quark distributions.
All the functions, $\f^i$, $g^i$ and $\theta^i_{\mp}$ can be
determined from extremization of the thermodynamic potential.

Realizing the fact that, the state given in Eq.(\ref{ubt}) is obtained by successive Bogoliubov
transformations, it is easy to calculate expectation values of different operators
in terms of the ansatz functions. In particular, the scalar condensate
for the i-th flavor can be written as

\bearr
\langle\Omega(\beta,\mu)|\bar\psi^i\psi^i|\Omega(\beta,\mu)\rangle
 &=& -\sum_{n=0}^\infty\frac{N_c|q^i|B\alpha_n}{(2\pi)^2}
\int{{dp_z}\cos\phi^i\cos 2g^i
\left(1-\sin^2\theta_-^i-\sin^2\theta_+^i\right)}\nonumber\\
&\equiv&-I_s^i
\label{isi}
\eearr
where, $\alpha_n=(2-\delta_{n,0})$ is the degeneracy factor of the
 $n$-th Landau level (all levels are doubly degenerate except the lowest Landau level).
As we shall see later, the functions $\sin^2\theta_\mp$ will be related to the
distribution functions for the quarks and anti quarks. Further,for later convenience, 
here we have introduced the function $\phi^i\equiv \phi_0^i-2 f^i$ with $\cot\phi_0^i=
m^i/\epsilon_{ni},\quad \epsilon^i=\sqrt{m^{i2}+2n |q^iB|}$ . Similarly, the pseudoscalar condensate is given as
\bearr
\langle\Omega(\beta,\mu)|\bar\psi^i\gamma_5\psi^i|\Omega(\beta,\mu)\rangle
 &=& -\sum_{n=0}^\infty\frac{N_c|q^i|B\alpha_n}{(2\pi)^2}
\int{{dp_z}\sin 2g^i
\left(1-\sin^2\theta_-^i-\sin^2\theta_+^i\right)}\nonumber\\
&\equiv&-I_p^i
\label{ipi}
\eearr
 Thus a non vanishing $I_s^i$ will imply a chiral symmetry breaking phase while
a non vanishing $I_p^i$ will indicate a CP violating phase.

 The energy density $\cal E$ can be calculated by taking the expectation
value of the Hamiltonian corresponding to the Lagrangian of Eq.(\ref{lag3fl})
with respect to the state given in Eq.(\ref{ubt}) as in Ref.s \cite{bhaswarnjlb,bhaswarcp}. 
The thermodynamic potential is then given by
\be
\Omega={\cal E} -\mu\rho -\frac{1}{\beta}s
\label{Omega}
\ee
In the above, $\mu $ is the quark chemical potential and 
$\rho$, the total number density of the quarks is given by
\be
\rho=\sum_{i=1}^3\rho^i=\sum_i \langle{\psi^i}^\dag\psi^i\rangle
 = \sum_{i=1}^3\sum_{n=0}^{\infty}\frac{N_c\alpha_n|q^iB|}{(2\pi)^2}
\int{dp_z\left[\sin^2\theta_-^i - \sin^2\theta_+^i\right]}.
\label{rho}
\ee
Finally, for the entropy density for the quarks we have \cite{tfd}
\begin{equation}
s = -\sum_i\sum_n\frac{N_c\alpha_n|q^i|B}{(2\pi)^2}\int{dp_z\lbrace
(\sin^2\theta_-^i\ln{\sin^2\theta_-^i} +
\cos^2\theta_-^i\ln{\cos^2\theta_-^i}) + (-\rightarrow +)\rbrace}.
\label{entropy}
\end{equation}

Now the functional  minimization of  the thermodynamic potential
 $\Omega$ with respect to  the chiral condensate function
$\phi ^i (p_z)$ and the pseudoscalar function $g^i(p_z)$leads to

\begin{eqnarray}
\tan \phi^i = \frac{|\zbf p^i|}{M_s^i}\;\;\;\;\;\;\;\;and\;\;\;\;\;\;\;\;
\tan{2g^i} = \frac{M_p^i}{\sqrt{{M_p^i}^2 + |\zbf p^i|^2}}.
\label{tanph}
\end{eqnarray}

  \noindent
 with $|\zbf p^i|=\sqrt{p_z^2+2 n |q^iB|}$ and $M_s^i$ and $M_p^i$ are respectively the scalar and pseudoscalar 
contributions to the total
mass for the $i-{th}$ flavor. They are given by the solutions of the coupled gap equations

\begin{eqnarray}
M_s^i &=& m^i + 4GI_s^i + K|\epsilon_{ijk}|\lbrace\cos\theta(I_s^j I_s^k - I_p^j I_p^k) - 
\sin\theta(I_s^j I_p^k + I_p^j I_s^k)\rbrace,
\label{scmass} \\
M_p^i &=& 4G I_p^i - K|\epsilon_{ijk}|\lbrace\cos\theta(I_s^j I_p^k + I_p^j I_s^k) - 
\sin\theta(I_p^j I_p^k - I_s^j I_s^k)\rbrace.
\label{psmass}
\end{eqnarray}

The above equations are actually {\em self consistent} equations for $M_s^i$ and $M_p^i$ because $I_s^i$ and $I_p^i$ are given
in terms of $M_s^i$ and $M_p^i$ as in Eq.s ( \ref{isi},\ref{ipi}) and Eq.(\ref{tanph}.
\noindent Finally, extremizing the thermodynamic potential with respect to the thermal function $\theta_{\mp}$
leads to

\be
\sin^2\theta_\pm^{i,n}=\frac{1}{\exp(\beta(\omega^{i,n}\pm\mu^i))+1},
\label{them}
\ee
where, $\omega_{i,n}=\sqrt{{M^i}^2+p_z^2+2n|q^i|B)}$ is the excitation energy with
the constituent quark mass $M^i=\sqrt{{M_s^i}^2+{M_p^i}^2}$ arising from both scalar and pseudoscalar condensates.

Substituting the solution for the
condensate function of Eq. (\ref{tanph}) and the thermal function given in
Eq.(\ref{them}) back in Eq.s (\ref{isi},\ref{ipi}) yields respectively the scalar and pseudoscalar condensates  as
\be
I^{i}_s=
 \sum_{n=0}^\infty\frac{N_c|q^i|B\alpha_n}{(2\pi)^2}
\int{dp_z}\Big (\frac{M_s^i}{\omega_{i,n}}\Big) \left(1-\sin^2\theta_-^i-\sin^2\theta_+^i\right).
\label{Iis}
\ee
\be
{I^i}^p=
 \sum_{n=0}^\infty\frac{N_c|q^i|B\alpha_n}{(2\pi)^2}
\int{dp_z}\Big (\frac{{M_p^i}}{\omega_{i,n}}\Big) \left(1-\sin^2\theta_-^i-\sin^2\theta_+^i\right).
\label{Iip}
\ee

Thus Eq.s(\ref{scmass}),(\ref{psmass}) and Eq.s(\ref{Iis}),(\ref{Iip}) define the self consistent mass
gap equation for the $i$-th quark flavor. Using the solutions for the condensate function
as well as the gap equations Eq.s(\ref{scmass},\ref{psmass}), the thermodynamic potential given
in Eq.(\ref{Omega}) reduces to

\begin{eqnarray}
\Omega &=& -\frac{N_c}{4\pi^2}\sum_{n=0}^{\infty}\alpha_n\sum_i|q^iB|\int{dp_z\omega^i_n}\nonumber\\
 &&- \frac{N_c}{4\pi^2\beta}\sum_{n=0}^{n_{max}}\sum_i|q^iB|\int{dp_z\left[\ln{\lbrace 1 + 
e^{-\beta(\omega^i_n-\mu)}\rbrace} + \ln{\lbrace 1+e^{-\beta(\omega^i_n+\mu^i)}\rbrace}\right]}\nonumber\\
 &&+
 2G_s\sum_i\left[{I_s^i}^2 + {I_p^i}^2\right] + 4K\left[\cos\theta\prod_{i=1}^3 I_s^i + 
\sin\theta\prod_{i=1}^3 I_p^i\right]
\nonumber \\
 &&- 2K|\epsilon_{ijk}|\left[\cos\theta I_p^i I_p^j I_s^k + \sin\theta I_s^i i_s^j I_p^k\right]
\nonumber \\
\label{thpotmag}
\end{eqnarray}

\noindent
In the above the first term is the zero temperature and zero density term in presence of a constant
magnetic field. the second term is the medium dependent term while the last two terms are the remaining interaction 
terms of the Lagrangian. 
For the CP violating parameter $\theta\rightarrow 0$, and the pseudoscalar density $I_p^i\rightarrow 0$, the
thermodynamic potential reduces to the same as in Ref.\cite{bhaswarnjlb}. 
The first term in Eq.(\ref{thpotmag}) is ultraviolet divergent which is also transmitted to the gap equations
 Eq.s (\ref{scmass}-\ref{psmass}) through the integrals $I_s^i$ and $I_p^i$ in Eq.s (\ref{Iis}--\ref{Iip}) and need to
be regularized to get any meaningful result.  There have been different regularization schemes to tackle this
divergence like 
Schwinger proper time method \cite{andreasrebhan,igormag}, a smooth cut off \cite{fukushimaplb}.  
  We perform the regularization as in Ref.\cite{providencia,bhaswarnjlb} by 
adding and subtracting a zero field (vacuum) contribution which is also divergent. This makes the first term
of Eq.(\ref{thpotmag}) a rather appealing form of separating the zero field vacuum contribution that is divergent,  
and,  a field dependent  contribution which is finite. The divergent zero field vacuum contribution is then evaluated
with a finite cutoff in the three momentum $\Lambda$ as is usually done in NJL model without magnetic field.
 Thus, we write the first term of 
Eq.(\ref{thpotmag}) as a sum of the vacuum contribution and the finite field contribution which 
is written in terms of Riemann-Hurwitz $\zeta$ function as \cite{bhaswarnjlb,providencia}

\begin{eqnarray}
&&-\sum_{i=1}^{3}\sum_{n=0}^{\infty}
\frac{N_c\alpha_n|q^iB|}{(2\pi)^2}\int{dp_z}\sqrt{{M^i}^2 + p_z^2 + 2n|q^i|B}
\nonumber\\
&=& -\frac{2N_c}{(2\pi)^3}\sum_{i=1}^3\int{d\vec p\sqrt{|\vec p|^2 + {M^i}^2}}
\nonumber\\
 &-& \frac{N_c}{2\pi^2}\sum_{i=1}^3|q^iB|^2\left[\zeta^\prime(-1,x^i) 
- \frac{1}{2}({x^i}^2-x^i)\ln{x^i} + \frac{{x^i}^2}{4}\right],
\label{t1}
\end{eqnarray}

\noindent
where, we have introduced the dimensionless quantity, $x^i = \frac{{M_s^i}^2 + {M_p^i}^2}{2|q^iB|} = 
\frac{{M^i}^2}{2|q^iB|}$, i.e. the mass parameter in units of the magnetic field and $\zeta^\prime(-1,x) 
= d\zeta(z,x)/dz|_{z=1}$ is the derivative of the Riemann-Hurwitz zeta function. The zero field 
vacuum term in Eq.(\ref{t1}) can be calculated using a sharp cutoff as it is usually done in NJL model.

Using Eq.(\ref{t1}), the scalar and pseudoscalar condensates as given in Eq.s(\ref{Iis}--\ref{Iip}) 
can also be separated into a (divergent) vacuum term, a 
field dependent term which is finite and a medium dependent term which is also finite. Thus we can 
write the scalar condensate as 

\begin{eqnarray}
I_s^i \equiv -\langle\bar\psi^i\psi^i\rangle &=& \frac{2N_c}{(2\pi)^3}\int_{|\zbf p|<\Lambda} 
{d\vec p}\frac{M_s^i}{\sqrt{\vec p^2 + {M^i}^2}}\nonumber\\
&+& \frac{N_c M_s^i|q^iB|}{(2\pi)^2}\left[x^i(1-\ln{x^i}) + \ln{\Gamma(x^i)} +
\frac{1}{2}\ln{\frac{x^i}{2\pi}}\right]\nonumber\\
&-&\sum_{n=0}^{n_{max}}\frac{N_c|q^i|B\alpha_n}{(2\pi)^2}
\int dp_z\frac{M_s^i}{\sqrt{{M^i}^2 + p_z^2 + 2n|q^iB|}}(\sin^2\theta_-^i +\sin^2\theta_+^i)
\nonumber\\
 &=& {I_s^i}_{vac} + {I_s^i}_{field}^{si}+{I_s^i}_{med}.
\label{Isif}
\end{eqnarray}

\noindent
The zero field vacuum contribution, $I_{vac}^{si}$, can be analytically calculated using a sharp 
momentum cutoff $\Lambda$ and can be written as

\begin{equation}
I_{s_{vac}}^{i} = \frac{N_c M_s^i}{2\pi^2}\left[\Lambda\sqrt{\Lambda^2+{M^i}^2}-{M^i}^2\log\left\lbrace
\frac{\Lambda+\sqrt{\Lambda^2+{M^i}^2}}{M^i}\right\rbrace\right].
\label{Isvac}
\end{equation}

\noindent
Further, since $|\zbf p|=\sqrt{p_z^2+2n|q^iB|}$, the condition of a sharp cutoff in magnitude of three momentum
leads  to a finite number of Landau levels that are filled up till $n=n_{max}$ which is given as
 $n_{max}=Int[\frac{\Lambda^2}{2|q^iB|}]$
when $p_z=0$. Further this condition also lead to a cutoff for  $|p_z|$ as $\Lambda^\prime=\sqrt{\Lambda^2-2n|q^i|B}$ for
a given value of the Landau level $n$.

Similarly, we can write the pseudoscalar condensate as 

\begin{eqnarray}
I_p^i \equiv i\langle\bar\psi^i\gamma_5\psi^i\rangle &=& \frac{2N_c}{(2\pi)^3}\int 
{d\vec p}\frac{M_p^i}{\sqrt{\vec p^2 + {M^i}^2}}\nonumber\\
&+& \frac{N_c M_p^i|q^iB|}{(2\pi)^2}\left[x^i(1-\ln{x^i}) + \ln{\Gamma(x^i)} +
\frac{1}{2}\ln{\frac{x^i}{2\pi}}\right]\nonumber\\
&-&\sum_{n=0}^{n_{max}}\frac{N_c|q^i|B\alpha_n}{(2\pi)^2}
\int dp_z\frac{M_p^i}{\sqrt{{M^i}^2 + p_z^2 + 2n|q^iB|}}(\sin^2\theta_-^i +\sin^2\theta_+^i)
\nonumber\\
 &=& {I_p^i}_{vac} + {I_p^i}_{field}+{I_p^i}_{med}.
\label{Ipif}
\end{eqnarray}

\noindent
Here also the zero field vacuum contribution, ${I_p^i}_{vac}$, can be analytically calculated 
using a sharp momentum cutoff $\Lambda$ and can be written as

\begin{equation}
{I_p^i}_{vac} = \frac{N_c M_p^i}{2\pi^2}\left[\Lambda\sqrt{\Lambda^2+{M^i}^2}-{M^i}^2\log\left\lbrace
\frac{\Lambda+\sqrt{\Lambda^2+{M^i}^2}}{M^i}\right\rbrace\right].
\label{Ipvac}
\end{equation}

\noindent
Now, using Eq.(\ref{t1}), the thermodynamic potential can be rewritten as 

\begin{eqnarray}
\Omega &=& \Omega_{vac} + \Omega_{field} + \Omega_{med} \nonumber\\
&&+ 2G_s\sum_i\left[{I_s^i}^2 + {I_p^i}^2\right] + 4K\left[\cos\theta\prod_{i=1}^3 I_s^i + 
\sin\theta\prod_{i=1}^3 I_p^i\right] \nonumber\\
&&- 2K|\epsilon_{ijk}|\left[\cos\theta I_p^i I_p^j I_s^k + \sin\theta I_s^i i_s^j I_p^k\right].
\label{thpf}
\end{eqnarray}

\noindent
In the above, $\Omega_{vac}$ is the vacuum contribution towards the thermodynamic potential and using a sharp 
cutoff, it can be analytically calculated as

\begin{eqnarray}
\Omega_{vac}&=&-2N_c\sum_i\int_{|\zbf p|<\Lambda}
\frac{d\zbf p}{(2\pi)^3}\sqrt{\zbf p^2+{M^i}^2}\nonumber\\
&\equiv& -\frac{N_c }{8\pi^2}\sum_i\left[(\Lambda^2+ {M^i}^2)^{1/2}(2\Lambda^2+ {M^i}^2)-
{M^i}^4\log\frac{\Lambda+\sqrt{\Lambda^2+{M^i}^2}}{ M^i}\right].
\label{omgvac}
\end{eqnarray}

\noindent
$\Omega_{field}$ is the field contribution to $\Omega$ and is given by

\begin{equation}
\Omega_{field}=
 - \frac{N_c}{2\pi^2}\sum_i|q^iB|^2\left[\zeta^\prime(-1,x^i) 
- \frac{1}{2}({x^i}^2-x^i)\ln{x^i} + \frac{{x^i}^2}{4}\right],
\label{omgfield}
\end{equation}

\noindent
where the derivative of the Riemann-Hurwitz zeta function $\zeta(z,x)$ at $z=-1$ is given by \cite{hurwitz}

\begin{equation}
\zeta^\prime(-1,x) =-\frac{1}{2} x\log x-\frac{1}{4} x^2+\frac{1}{2} x^2\log x+
\frac{1}{12}\log x+x^2 \int_0^\infty \frac{2\tan^{-1} y+y \log(1+y^2)}{\exp(2\pi x y)-1} dy.
\label{rhd}
\end{equation}

\noindent
Finally, the  medium contribution $\Omega_{med}$ towards the thermodynamic potential is given by

\begin{equation}
\Omega_{med} =- \sum_{n,i}\frac{N_c\alpha_n|q^iB|}{(2\pi)^2\beta}\int{dp_z\left[\ln{\left\lbrace 1 
+ e^{-\beta(\omega^i_n-\mu)}\right\rbrace} + \ln{\left\lbrace 1+e^{-\beta(\omega^i_n+\mu)}
\right\rbrace}\right]}.
\label{omgmed}
\end{equation}

The coupled mass gap equations Eq.(\ref{scmass}), Eq.\ref{psmass}) and the thermodynamic potential Eq.(\ref{thpf}) constitute the
basis for our numerical results for various physical situations that we discuss in the following section.
\section{Results and discussions}

The three flavor NJL model that we investigate here, has five parameters in total, 
namely the current 
quark masses for the non strange and strange quarks, $m_q$ and $m_s$, the two couplings $G_s$, $K$ 
and the three-momentum cutoff $\Lambda$. We have chosen here $\Lambda=0.6023$ GeV, 
$G_s\Lambda^2=1.835$, $K\Lambda^5=12.36$, $m_u=5.5$ MeV $m_d$ and $m_s=0.1407$ GeV as has been used in 
Ref.\cite{rehberg}. After choosing $m^q=5.5$ MeV, the remaining four parameters are fixed by fitting 
to the pion decay constant and the masses of pion, kaon and $\eta'$. With this set of parameters the 
mass of $\eta$ is underestimated by about six percent and the constituent masses of the light quarks 
turn out to be $M^{u,d}=0.368$ GeV for u-d quarks, while the same for strange quark turns out as 
$M^s=0.549$ GeV, at zero temperature and zero density.

In the numerical calculations that follows, we have taken the quark chemical potential 
$\mu$ to be same for all the three flavors. For a given values of $T$, $\mu$ and strength of magnetic field $B$,
 we first solve the gap equations (\ref{scmass}) and (\ref{psmass}) self consistently along with the condensates given
in Eq.(\ref{Isif}) and Eq.(\ref{Ipif}) with the parameters of the NJL model as above. 
Since we have assumed $m_u=m_d$, the equations actually represent four coupled equations for zero 
magnetic field : two corresponding to the scalar contributions towards the masses, i.e, $M_s^u=M_s^d$ 
and $M_s^s$ and two corresponding to the pseudoscalar contributions towards the masses, i.e, 
$M_p^u=M_p^d$ and $M_p^s$. However, this degeneracy is lifted in presence of finite magnetic field. Thus
Eq.s (\ref{scmass}) and (\ref{psmass}) actually represent six coupled mass gap equations-- the contributions to the
masses arising from the scalar and pseudoscalar condensates of each flavor.
Once the solutions to these coupled equations for the masses and the condensates are found, they are 
then substituted in Eq.(\ref{thpf}) to find the thermodynamic potential $\Omega$. In 
case of more than one solution to the gap equation, the solution with the minimum $\Omega$ is 
chosen.

In our analysis, we have explored the behavior of scalar and pseudoscalar contributions to the quark 
mass with temperature, chemical potential and magnetic field for different values of $\theta$. 

\begin{figure}[h]
\vspace{-0.4cm}
\begin{center}
\begin{tabular}{c c c}
\includegraphics[width=6cm,height=6cm]{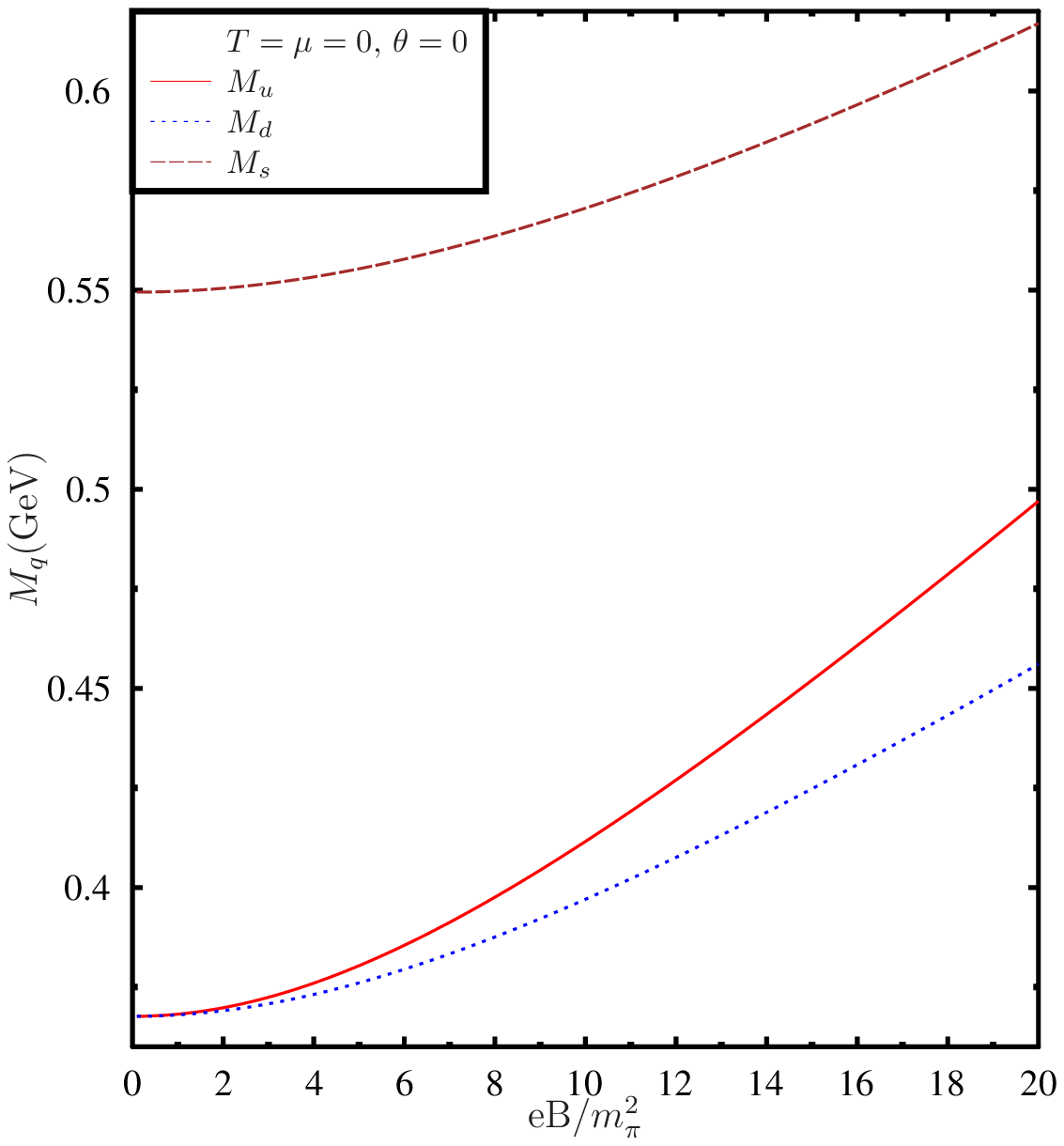}&
\includegraphics[width=6cm,height=6cm]{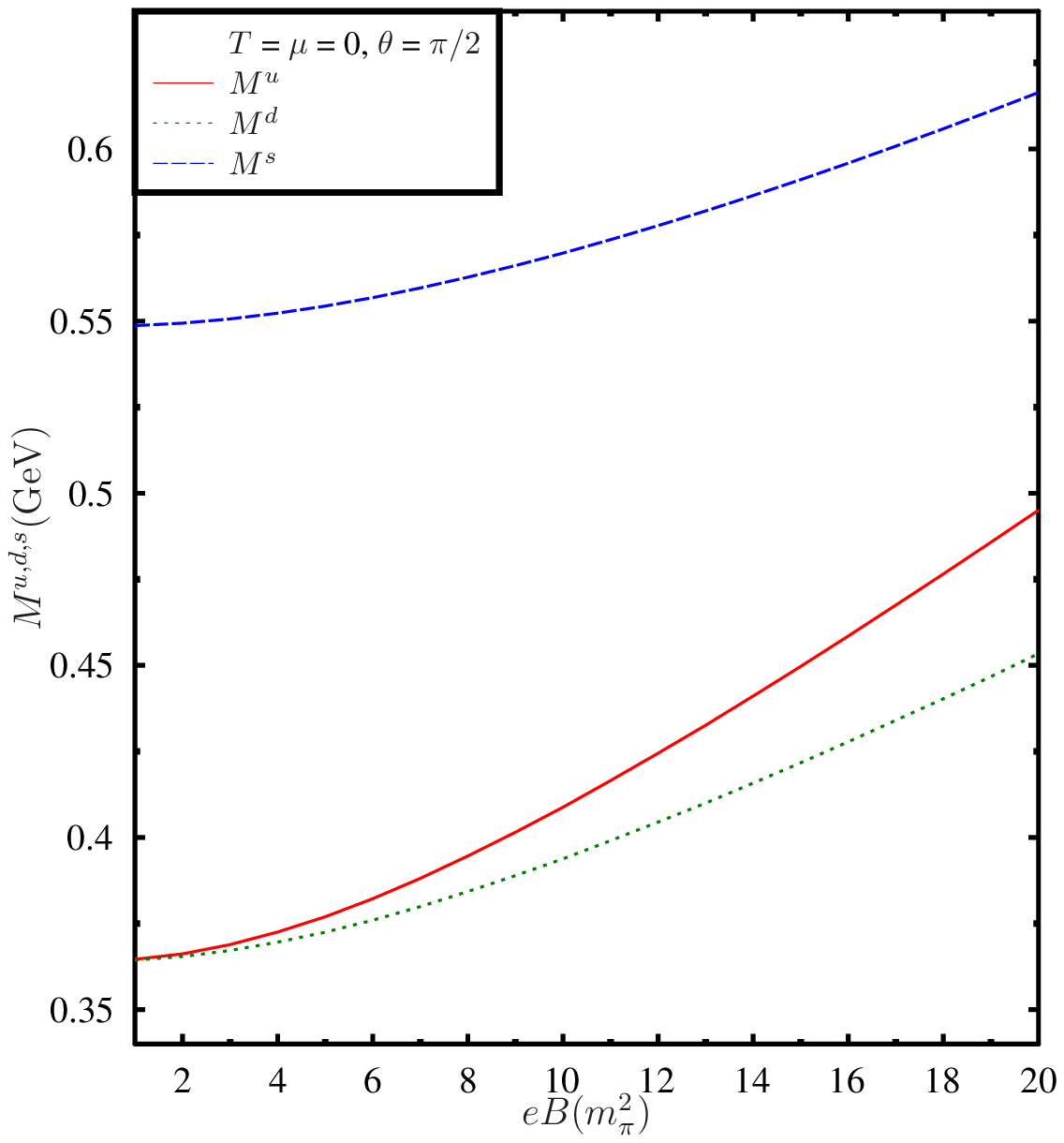}&
\includegraphics[width=6cm,height=6cm]{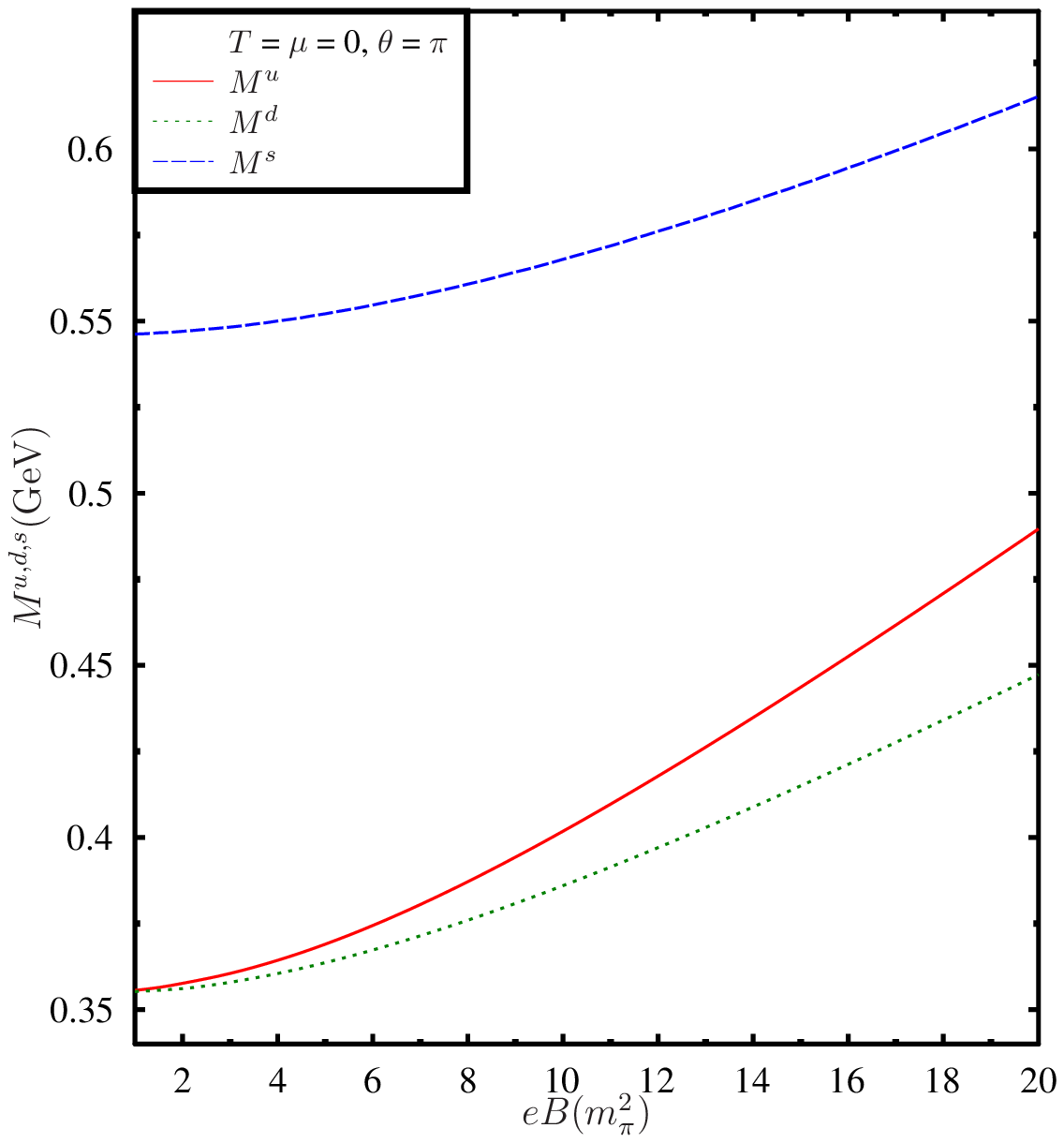}\\
Fig. 1-a & Fig. 1-b & Fig. 1-c
\end{tabular}
\end{center}
\caption{Constituent quark mass at T=0,$\mu=0$ as a function of magnetic field
for  $u$- quark for $\theta=0$(Fig. 1-a), $\theta=\pi/2$(Fig. 1-b) and $\theta=\pi$(Fig.1-c)}
\label{fig1}
\end{figure}

Let us first discuss the effect of magnetic field on the ``vacuum" properties within the model.
In Fig.\ref{fig1} we have plotted the constituent quark masses given as $M^i=\sqrt{M_s^{i2}+M_p^{i2}}$ 
as a function of magnetic field for different representative values of $\theta$. 
Due to the different charges of the u and d quarks, the isospin symmetry is lost between the light quarks when
an external magnetic field is applied to the system. 
The magnetic catalysis of dynamical generation of mass is seen for all the quarks with the constituent quark
masses increasing with magnetic field for all values of $\theta$. The constituent quark masses here, however, are
 generated from from quark anti quark condensates both 
in scalar and pseudoscalar channels  for non zero values of $\theta$.

\begin{figure}
\includegraphics[width=8cm,height=8cm]{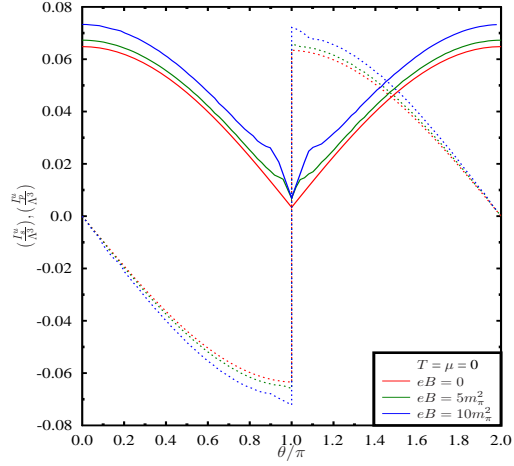}
\caption{ $\theta$ dependence of the order parameters for u -quarks  for different magnetic fields. The solid lines
correspond to condensates in the scalar channel while the dotted lines 
correspond to condensates in the pseudoscalar channel.
$\theta=\pi/2$.}
\label{fig2}
\end{figure}

In Fig.\ref{fig2}  the condensates in 
scalar and pseudoscalar channel for u-quark is plotted  as $\theta$ increases for different magnetic fields.
The condensates in the scalar and pseudoscalar channel vary in a complimentary manner so that the 
total constituent mass remains almost constant as $\theta$ is varied. This behavior is also seen with increasing magnetic 
field along with the fact that
the condensates in both the channels become larger in magnitude  for larger magnetic field.
The spontaneous CP violation is seen for $\theta=\pi$ with two degenerate solution for the 
pseudoscalar condensate differing by a sign.

\begin{figure}
\includegraphics[width=8cm,height=8cm]{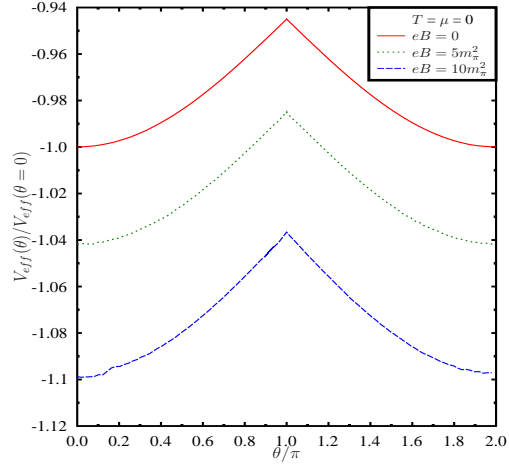}
\caption{Effective potential at $T = \mu = 0$ as a function of $\theta$ for different
strengths of magnetic field..}
\label{fig3}
\end{figure}

Fig.\ref{fig3} shows the variation of the effective potential with $\theta$ for different strengths of
magnetic fields. The effective potential shown here is normalized with respect to the effective 
potential at $\theta = 0$. It is minimum when $\theta = 0$ which is consistent with the Vafa-Witten 
theorem. The behavior we see here is similar to what we observed without the magnetic field. The 
magnetic field only reduces the effective potential.

\begin{figure}[h]
\vspace{-0.4cm}
\begin{center}
\begin{tabular}{c c c}
\includegraphics[width=6cm,height=6cm]{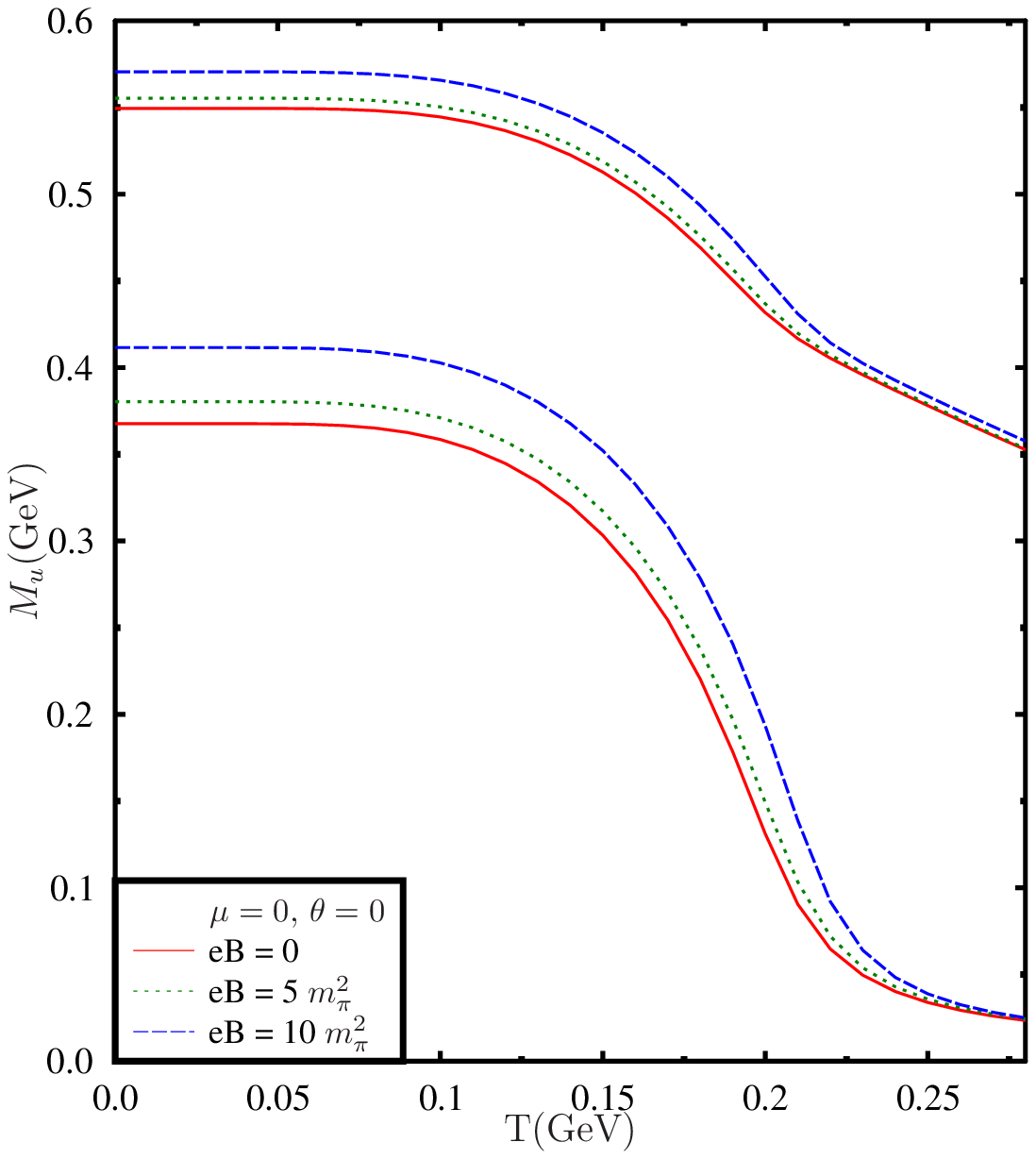}&
\includegraphics[width=6cm,height=6cm]{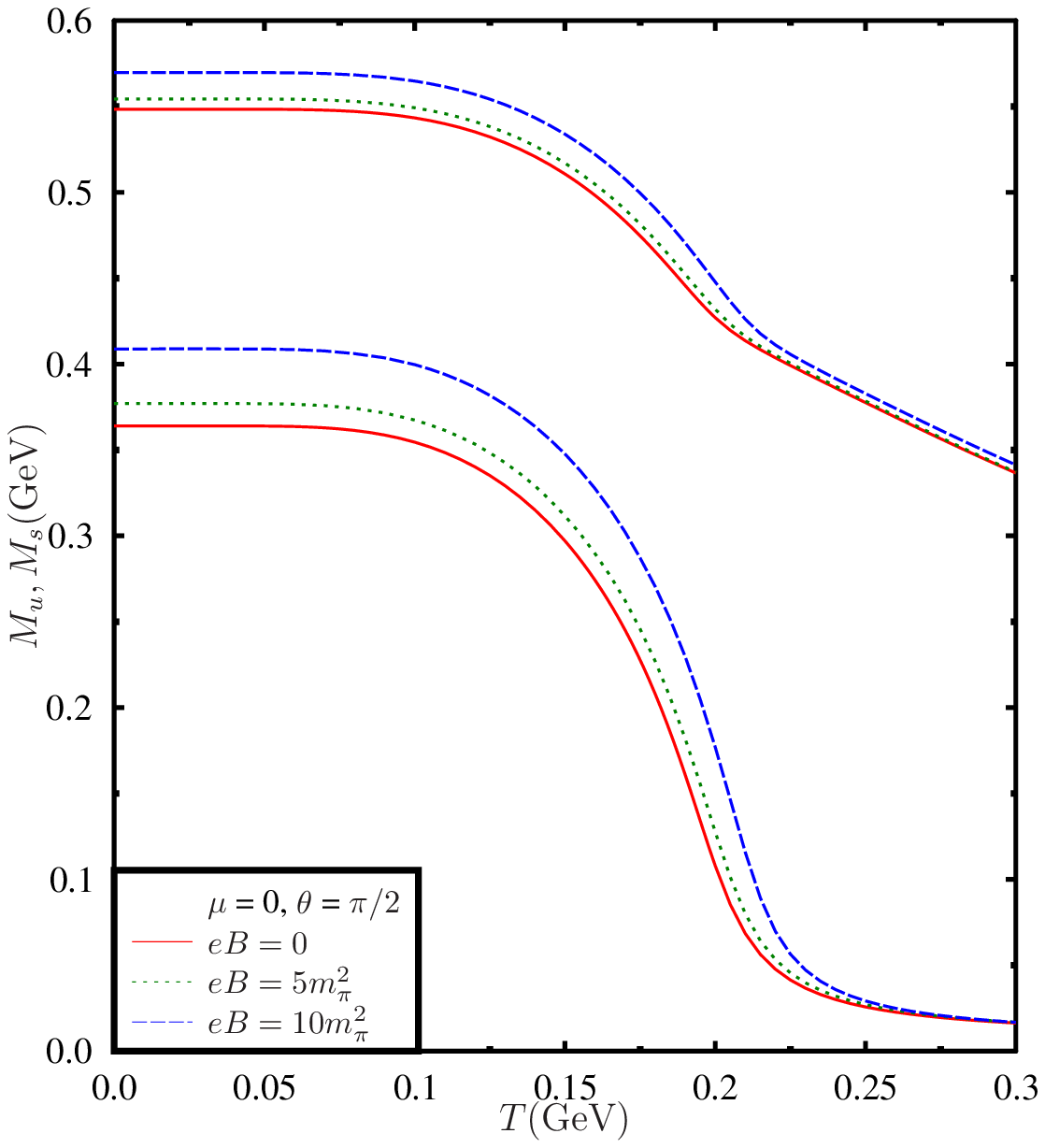}&
\includegraphics[width=6cm,height=6cm]{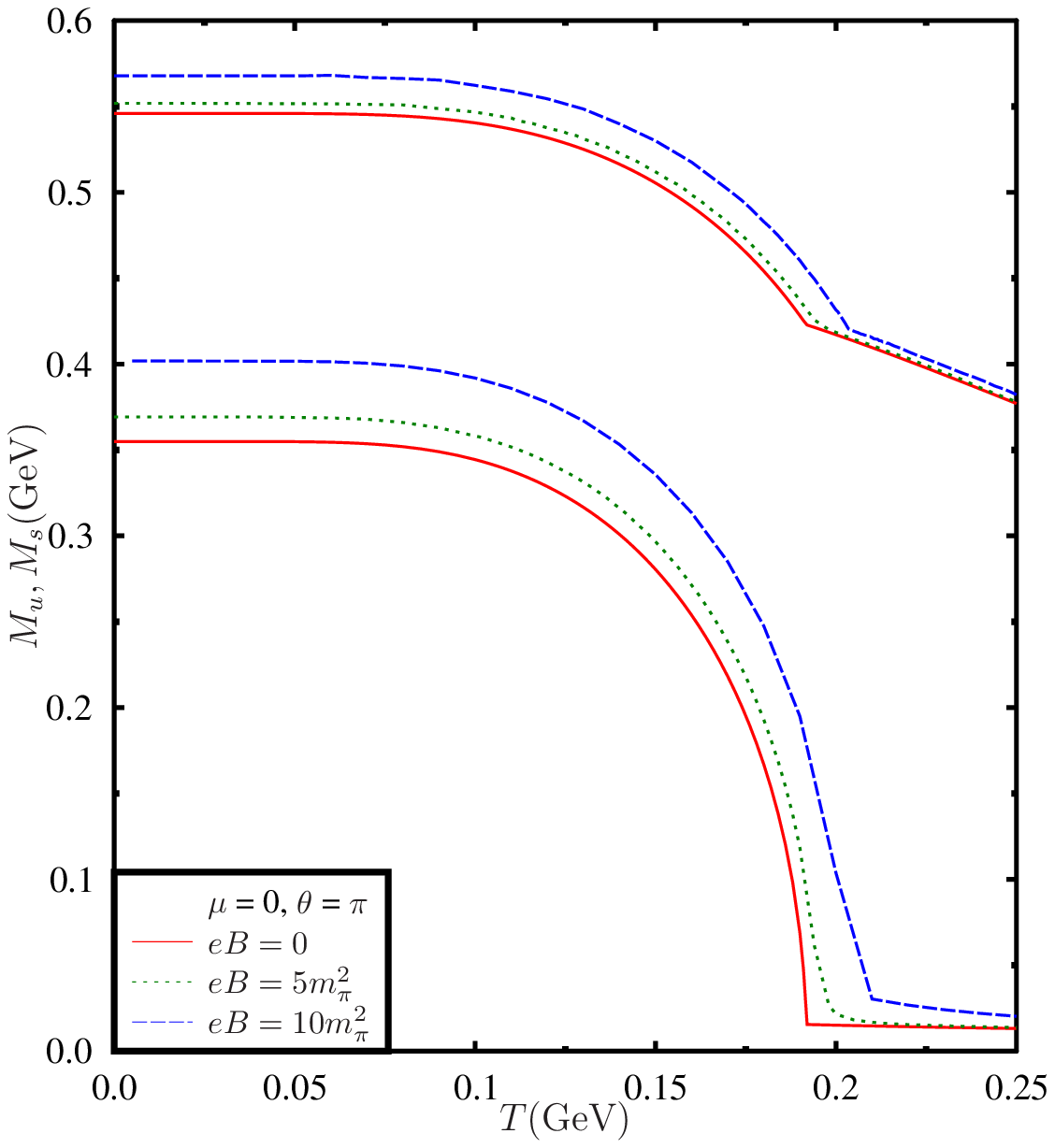}\\
Fig. 4-a & Fig. 4-b & Fig. 4-c
\end{tabular}
\end{center}
\caption{Temperature dependence of constituent quark masses for up and strange quarks 
for $\theta=0$(Fig. 4-a), $\theta=\pi/2$(Fig. 4-b) and $\theta=\pi$(Fig.4-c) with different
strengths of magnetic field. In each plot the lower curves are for the up quark mass variation while the upper curve
shows temperature dependence of the strange quark mass .}
\label{fig4}
\end{figure}

In Fig.(\ref{fig4}) we show the temperature dependence  of the total mass for the u- quark for different magnetic fields
at zero chemical potential. For vanishing $\theta$, the total mass gets contribution only from the condensates in the 
scalar channel. Within the model, the chiral transition temperature increases with the 
magnetic field 
similar to several effective models as well as some lattice QCD models \cite{mukhlb}.
The  general reason being, magnetic field enhances the condensates and hence requires higher temperatures 
to melt the condensate.
As a result of the charge difference we obtain
a higher transition temperature for u-quark than for d-quark with the difference 
becoming larger with larger magnetic fields. 
The chiral transition is a crossover due to finite current quark mass. 
 However, in some of the recent
lattice calculations, inverse magnetic catalysis near the critical
temperature is observed leading to a 
to reduction of the crossover 
transition temperature with magnetic field \cite{schaferlat}. At  sufficiently
lower temperature, on the other hand, magnetic catalysis is observed in these lattice simulations with the
condensates getting enhanced with magnetic field. Such an effect can be generated in an ad hoc manner by 
reducing the effective four fermion coupling by making it a function of temperature and magnetic field
as in Ref.\cite{ferreira}.  There have been other attempts to
explain this by invoking paramagnetic contributions to the pressure with
large magnetization\cite{noronha}; magnetic inhibition due to neutral meson
fluctuation \cite{fukuprl} as well as a back reaction of the Polyakov loop 
which could be affected by magnetic field \cite{endrodi}. In the present work
however, we shall continue to consider the consequences of the ansatz
as in Eq.(\ref{ansatz}), to discuss the effect of  the non vanishing $\theta$
and magnetic field on the phase structure within the premises of
NJL model.

 As $\theta$ is increased, the contribution to the mass from 
the pseudoscalar condensates  also increases. We have also plotted the temperature dependence of the  pseudoscalar  component
of u-quark mass arising from pseudoscalar condensates $M_u^p$
for $\theta=\frac{\pi}{2}$ and $\theta=\pi$ in Fig. 5. 
As may be observed from Fig. 5, for $\theta=\pi/2$ the CP transition is a 
crossover transition.  AT $\theta=\pi$, however, the CP transition is a second order transition 
with the pseudoscalar condensates
smoothly vanishing at the critical temperature. Further, this CP restoration transition 
temperature increases with magnetic field. We should however note that what we considered here is the
equilibrium uniform CP violating phase structure induced by the determinant term. However,
local parity violating phase can also arise due to fluctuations of topological charges
induced through sphaleron configuration which are not exponentially suppressed \cite{dimacp}. On the other hand
such domains can also arise due to non equilibrium situations depending upon the
kinetics of the phase transition. 
Such CP odd
domains can decay via CP odd processes and can have observable effects like
chiral magnetic effect for non central heavy ion collisions\cite{cme} as well as
possible excess in dilepton production for central collisions \cite{andrianovplb}.
 
\begin{figure}[h]
\vspace{-0.4cm}
\begin{center}
\begin{tabular}{c  c}
\includegraphics[width=6cm,height=6cm]{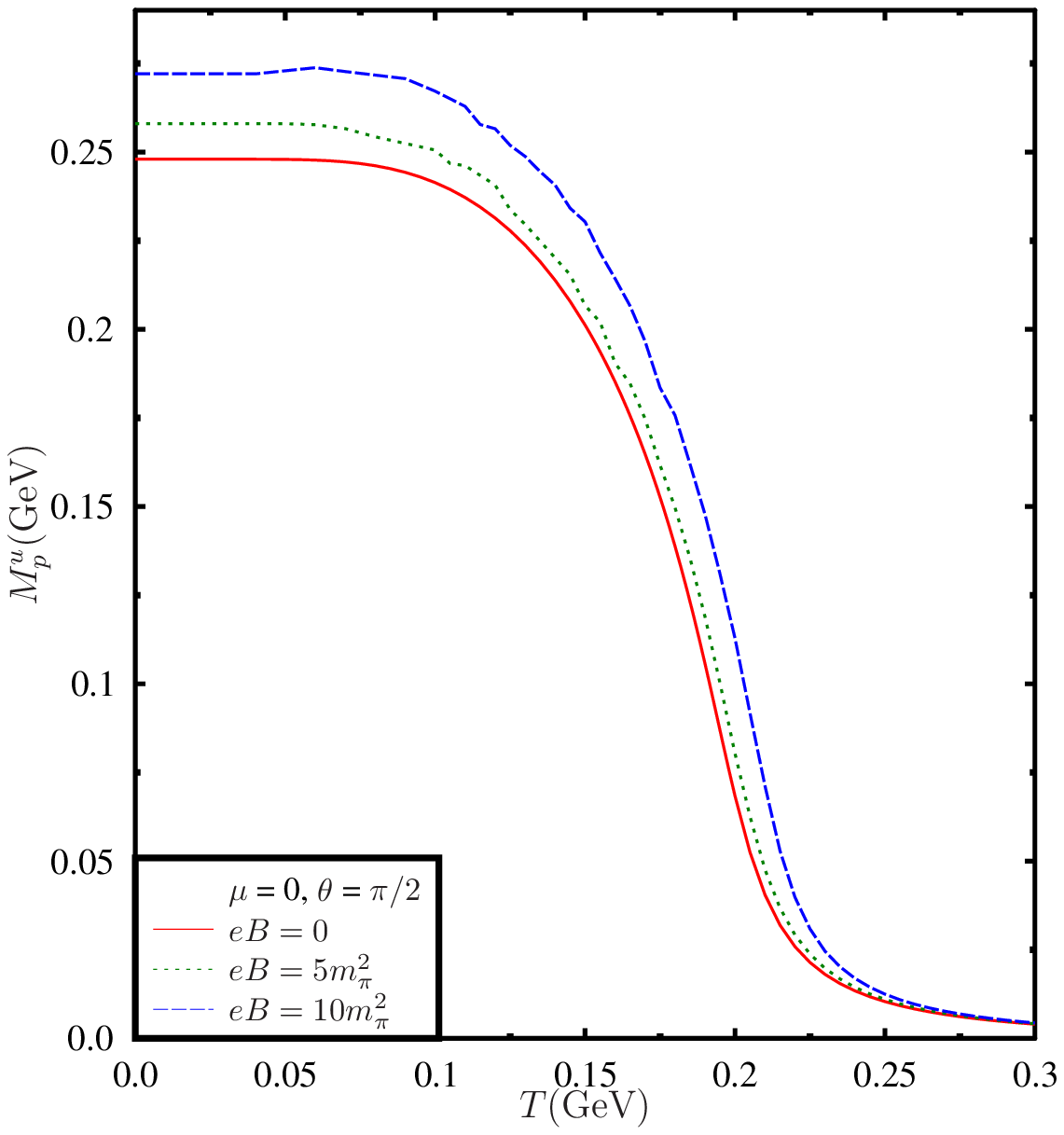}&
\includegraphics[width=6cm,height=6cm]{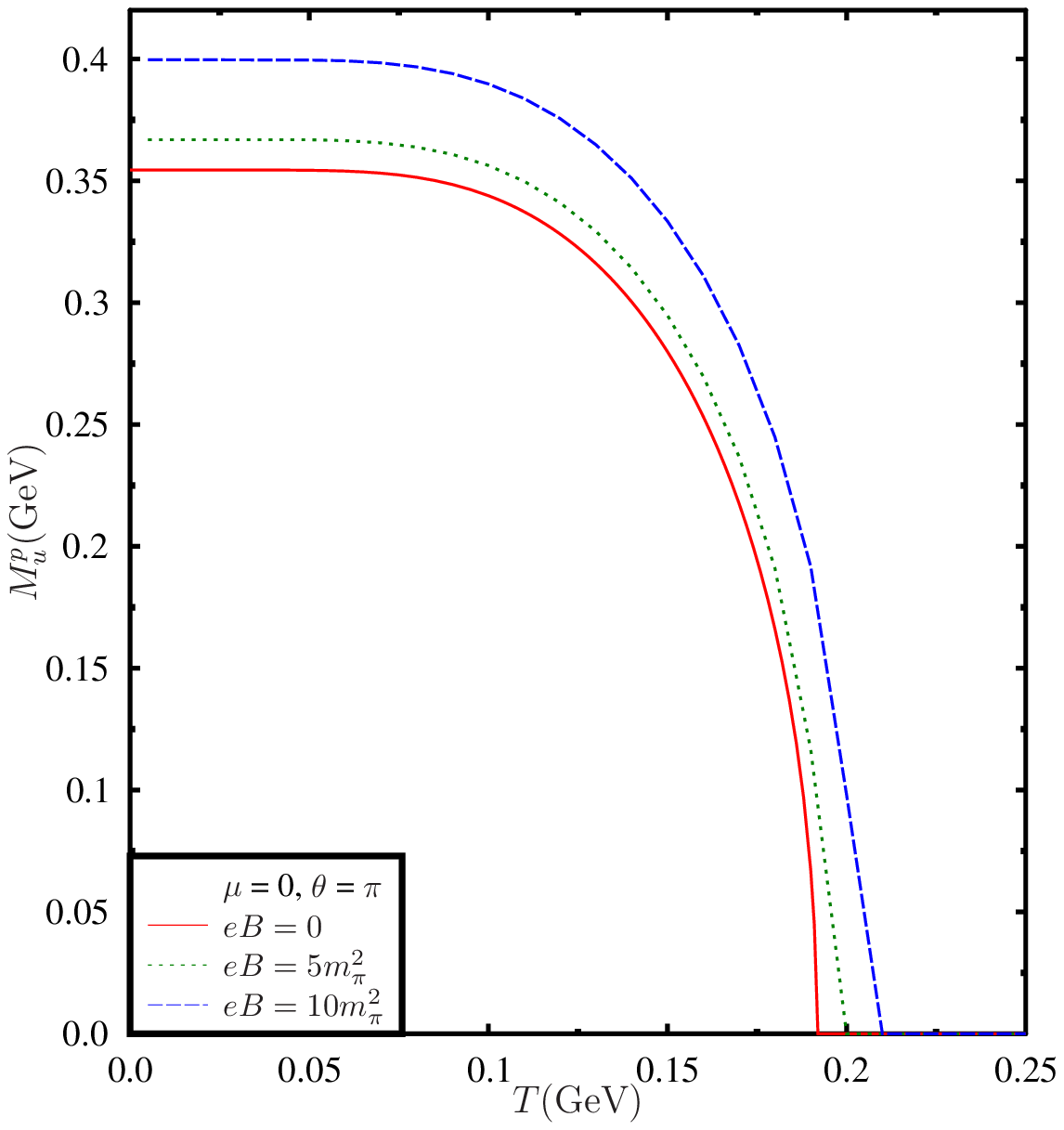}\\
Fig. 5-a & Fig. 5-b 
\end{tabular}
\end{center}
\caption{Temperature dependence of the pseudoscalar condensates
for $\theta=\pi/2$(Fig. 5-a) and $\theta=\pi$(Fig.5-b) with different
strengths of magnetic field. 
}
\label{fig5}
\end{figure}


\begin{figure}[h]
\vspace{-0.4cm}
\begin{center}
\begin{tabular}{c c c}
\includegraphics[width=6cm,height=6cm]{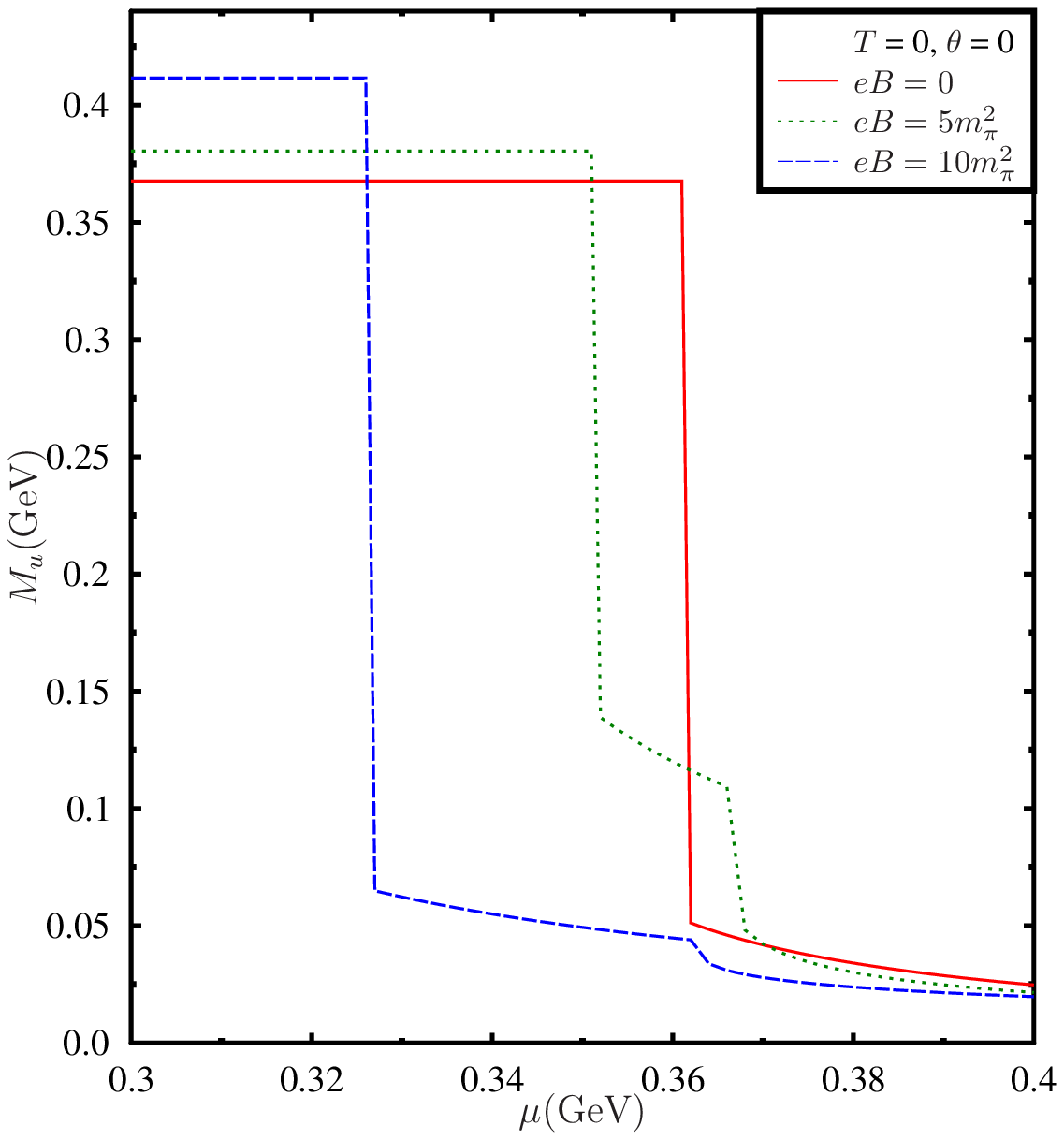}&
\includegraphics[width=6cm,height=6cm]{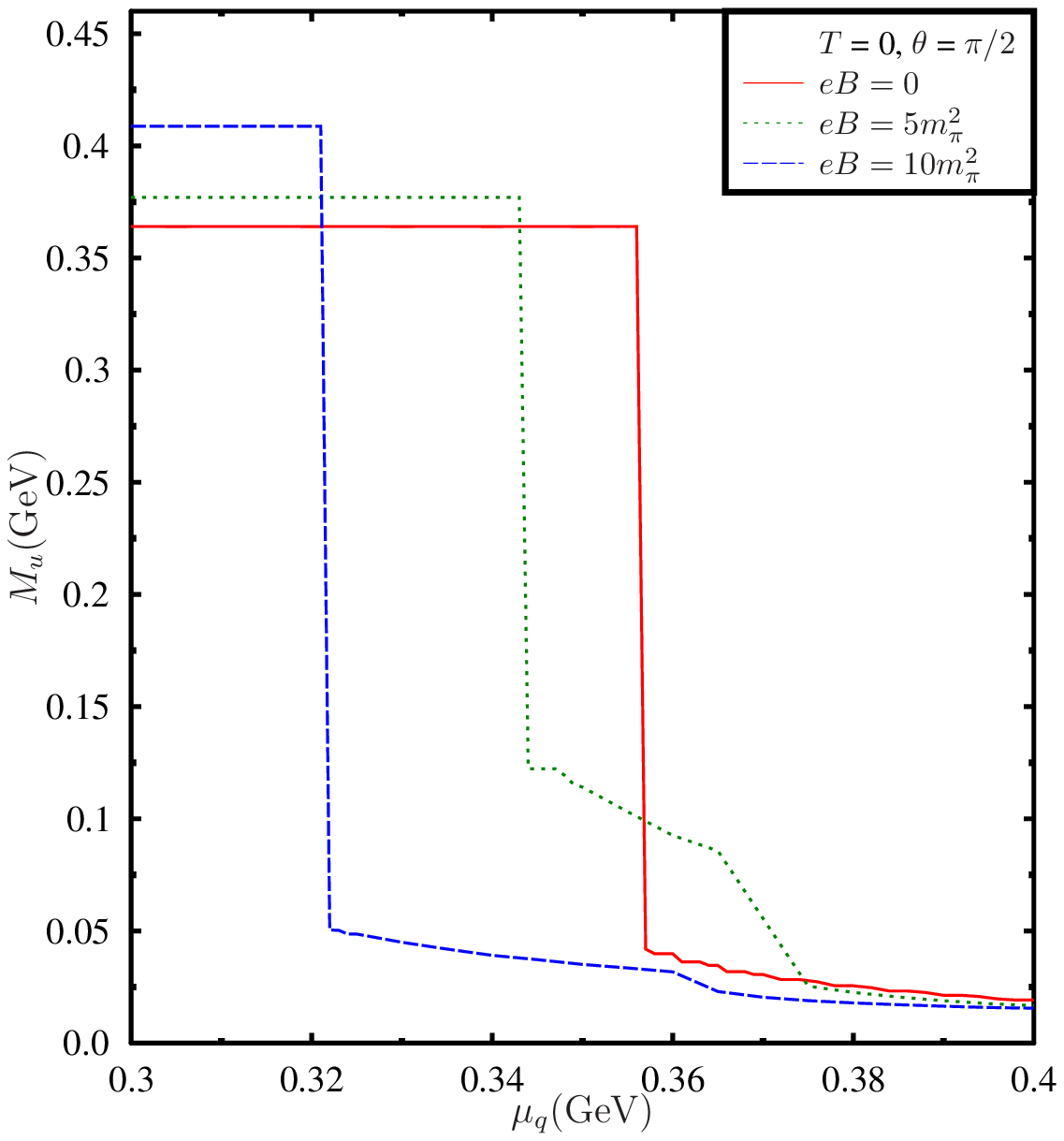}&
\includegraphics[width=6cm,height=6cm]{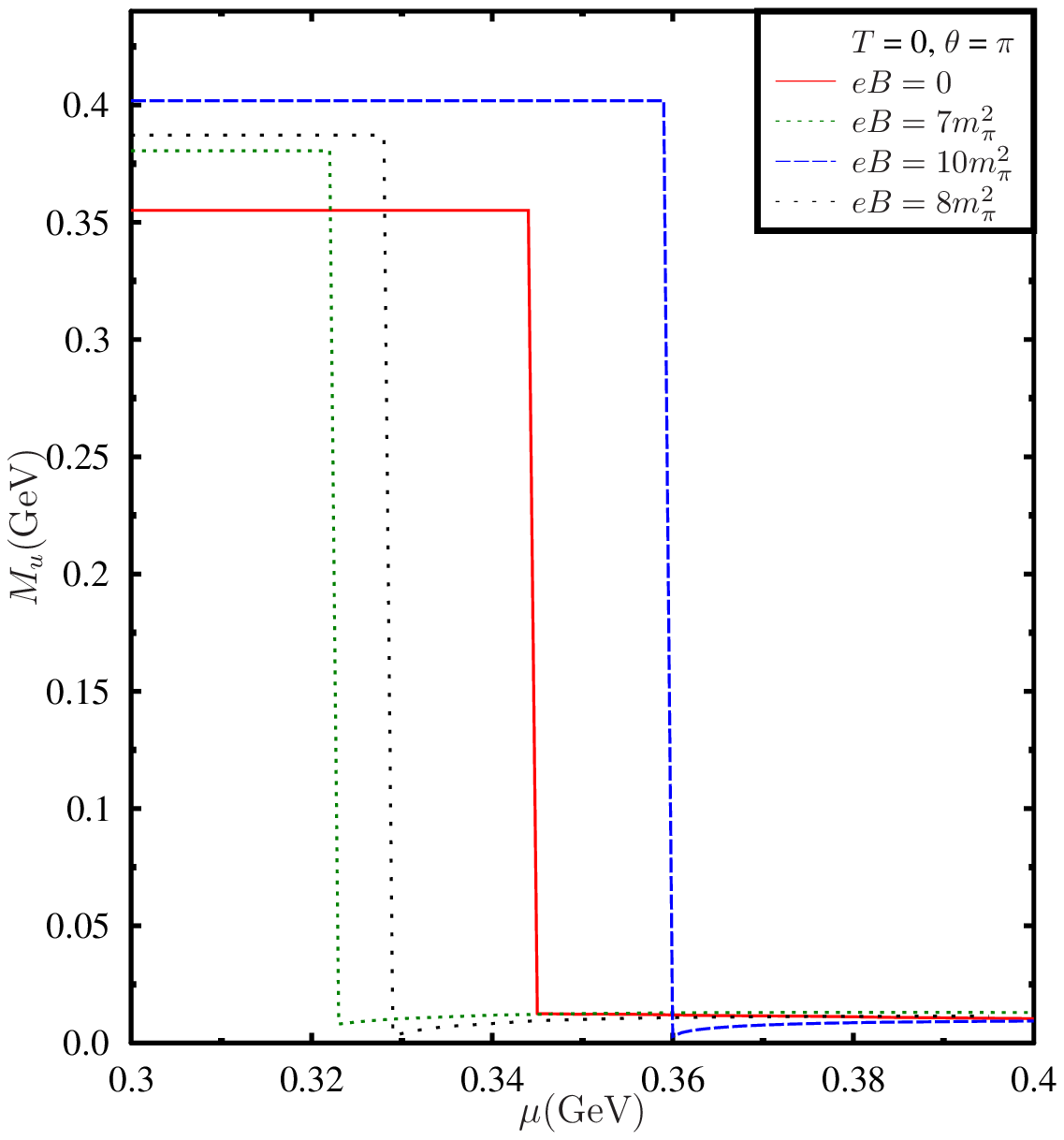}\\
Fig. 6-a & Fig. 6-b & Fig. 6-c
\end{tabular}
\end{center}
\caption{ Up quark mass as a function of quark chemical potential at zero temperature 
for $\theta=0$(Fig. 6-a), $\theta=\pi/2$(Fig. 6-b) and $\theta=\pi$(Fig.6-c) with different
strengths of magnetic field.
}
\label{fig6}
\end{figure}

In Fig(\ref{fig6}), we display the dependence of the constituent quark mass on quark chemical potential at zero temperature.
The critical chemical potential where the total quark mass shows a discontinuity decreases with magnetic field. The transition
is a first order one at zero temperature. Let us note that for $\theta=0$, the entire mass arises from
quark condensates in the scalar channel 
apart from the current quark masses. This behavior of having lower critical chemical potential for higher
magnetic field is the phenomenon of inverse magnetic catalysis of chiral symmetry breaking at finite chemical potential
\cite{andreasrebhan,bhaswarnjlb}. For finite $\theta$ however,
the mass is generated by condensates in both scalar and pseudoscalar channels. It turns out that for $\theta=\pi$, 
at zero magnetic field the critical quark chemical potential is $\mu_c\sim 545$ MeV with a first order transition .
As magnetic field is increased, $\mu_c$ decreases and is minimum at $eB=7 m_\pi^2$ with $\mu_c\sim 523$MeV. As the magnetic 
field is increased further, the the critical 
chemical potential also  increases and becomes about $\mu_c\sim 560$ MeV for $eB=10 m_\pi^2$. Such a behavior of decrease 
of critical chemical potential for intermediate strengths of magnetic field and then increase for stronger fields 
is also observed in Ref. \cite{florian}.
This inverse magnetic catalysis of CP transition at finite $\theta$ can be understood in a manner similar 
to Ref.\cite{florian}
discussed for chiral symmetry breaking in strong interaction. This can be easily done by  analyzing the pseudoscalar mass gap
equation Eq.(\ref{psmass}). The analysis can be easily carried out for $\theta=\pi$ for which we can approximate $I_s^i\simeq 0$
as well as $M_s^i\simeq 0$ for the light quarks. For large coupling, the solution to the gap equation
is given by  the $\mu=0$ solution.  For nonzero but small magnetic fields,
we can get the solution up to second order in magnetic field as
\be
M_p^i\simeq M_0^i\left(1+\frac{G_s|q_iB|^2}{M_0^{i2}(1-\frac{6G}{\pi^2}f(M_0^i,\Lambda))}\right),
\label{mpiapprox}
\ee
where, $f(M_0^i,\Lambda)=\Lambda^2\left[(1+{\hat M}_0^{i^2})-
{\hat M}_0^{i2}\log(\frac{1+{\hat M}_0^{i2}/2}{{\hat M}_0^i})\right]$;,
${\hat M_0^i}=M_0^i/\Lambda$ and, $M_0^i$ is the solution of the pseudoscalar mass gap equation Eq.(\ref{psmass}) with
$\mu=0=B$. One can substitute this solution in the thermodynamic potential and subtract out $M_p^i=0$ free energy density. 
In the limit of nonzero but small magnetic field, the difference in thermodynamic potential
is given as
\begin{equation}
\Delta \Omega\simeq -\frac{3}{8\pi^2}\sum_i M_0^{i2}\Lambda^2 \left(1-\frac{\pi^2}{3 G\Lambda^2}\right)+
\frac{3}{4\pi^2}|q_iB|\mu^2
-\frac{3}{4\pi^2}|q_iB|^2\left(1+\log\frac{M_0^{i2}}{2|q_i|B}\right)
\end{equation}
In the above, the first term is the vacuum (T=0, B=0,$\mu=0$) contribution to the free energy difference. The term linear in the magnetic field correspond to the free energy cost to form a quark anti quark pair in the pseudoscalar channel at finite $\mu$ which,
also depends upon the magnetic field along with the chemical potential.
 The last term is the gain in thermodynamic potential due to condensation which is
quadratic in magnetic field strength. Therefore, as we turn on the magnetic field, and start from broken phase with $\Delta\Omega<0$, for small field it can make $\Delta\Omega$ positive with the symmetry restored. However, as the field strength is increased,
the quadratic term starts dominating and symmetry broken phase is preferred again and thus the critical chemical potential
 will increase with magnetic field. The behavior of the critical chemical potential with magnetic field is reflected  
in Fig.\ref{fig6} for $\theta=\pi$ case.

\section{Summary} In the present work, we have focussed on the effect of $\theta-$ vacuum on the chiral transition for
hot and dense matter in the presence of magnetic field. The effect of CP violating $\theta-$ term in QCD in incorporated through
a $\theta$ dependent flavor mixing determinant interaction within a 3- flavor NJL model Lagrangian. The
methodology uses an explicit variation construct for the ground state in terms of quark anti-quark paring, 
instead of performing a chiral rotation of quark fields\cite{sakai,kuonop}. The ansatz function in the variational construct for the ground
state are determined from the minimization of the thermodynamic potential solving self consistent gap equations for the 
condensates in the scalar as well as the pseudoscalar condensates.

For non vanishing $\theta$, the constituent quark masses arise from quark anti quark condensates both in scalar and pseudoscalar
channels. With increasing values of CP violating parameter $\theta$, the pseudoscalar condensates increase
  and become maximum in magnitude at $\theta=\pi$. On the other hand, while the condensate
in the scalar channel decrease with $\theta$ and almost vanish for $\theta=\pi$ but for the current quark mass contribution.
The condensate in the two channels vary in a complimentary way such that constituent quark mass remains almost constant with
$\theta$ variation. Magnetic field enhances the condensates in {\em both}  the channels and breaks the isospin symmetry of
the light quarks.

The effective potential as a function of $\theta$ shows the minimum at $\theta=0$ with cusp at $\theta=\pi$ consistent with
the Vafa-Witten theorem. Introduction of magnetic field does not change this behavior. It only reduces the magnitude of 
the effective potential.

 At vanishing chemical potential, with temperature, the condensates in both the channel decrease. The CP transition
is a second order transition at around $T_c=200 MeV$. With magnetic field, this CP transition still remains
second order with the transition  temperature increasing as magnetic field strength is increased similar to the
magnetic catalysis of chiral symmetry breaking at finite temperature. This high temperature restoration of CP is expected as
the instanton effects responsible for CP violating phase become suppressed
exponentially at high temperatures \cite{grosspisarski}.

At finite chemical potential however, the CP transition is a first order transition. Further, inverse magnetic
catalysis for the CP transition is observed at finite chemical potential at zero temperature i.e.
the corresponding critical chemical potential decrease with magnetic fields for  small magnetic fields.
Possibility of a first order phase transition can lead to formation of CP-odd
meta stable domains which could be of relevance for heavy ion collisions at the Facility for Anti proton and Ion Research (FAIR)
as well as at Nucleotron based Ion Collider facility (NICA) at Dubna. However, it ought to be mentioned that
for the application to heavy ion collision, it is crucial to include
the non equilibrium dynamics of formation of domains, which will provide the relevant time scales and also provide
informations on possibility
of measuring the effects arising from the formation of such CP odd domains\cite{hmsanjay}.

We have considered here quark-anti quark pairing in our ansatz
for the ground state which is homogeneous with zero total momentum
as in Eq.(\ref{ansatz}). However, it is possible that
the condensate could be spatially non-homogeneous with a net total
momentum \cite{dune,frolov,nickel} or for very strong fields
could be non isotropic with vector condensation\cite{chernodub}. Further, one could  
include the effect of
deconfinement transition by generalizing the present model to Polyakov
loop NJL models for three flavors to investigate the inter relationship
of deconfinement and the chiral transition\cite{gato} as well as CP violation\cite{kuonop} in 
presence of strong fields
for  the three flavor case
considered here. This will be particularly important
for finite temperature  and low baryon densities. On the other hand,
at finite density and small
temperatures, the ansatz can be generalized to include the
diquark condensates in presence of magnetic field \cite{digal,iran,ferrerscmag}.
Some of these calculations are
in progress and will be reported elsewhere.
\begin{acknowledgments}
One of the authors (AM) would like to acknowledge Department of Science and Technology, Government of
India ( project no. SR/S2/HEP-031/2010) for financial support.
\end{acknowledgments}

\def\endm{C. Baker et al., {\PRL{97}{131801}{2006}};
 J. Kim and G. Carosi, Rev. Mod. Phys. {\bf 82}, 557 (2010).}
\def\pecceiquinn{R. D. Peccei and H. R. Quinn, {\PRL{38}{1440}{1977}}; {\PRD{16}{1791}{1977}}.}
\def\vafawit{C. Vafa and E. Witten, {\PRL{53}{535}{1984}}.}
\def\dashen{R. Dashen, {\PRD{3}{1879}{1971}}}
\def\chpt{P. Vecchia and G. Veneziano, {\NPB{171}{253}{1980}};
 A. Smilga, {\PRD{59}{114021}{1999}}; M. Tytgat, {\PRD{61}{114009}{2000}};
 G. Akemann, J. Lenaghan and K. Splittorff, {\PRD{65}{085015}{2002}};
 M. Creutz, {\PRL{92}{201601}{2004}};
 M. Metlitski and A. Zhitnitsky, {\NPB{731}{309}{2005}}; {\PLB{633}{721}{2006}}.}
\def\fragalsm{A. Mizher and E. Fraga, {\NPA{820}{247c}{2009}}; {\NPA{831}{91}{2009}}.}
\def\cpnjl{T. Fujihara, T. Inagaki and D. Kimura, {\PTP{117}{139}{2007}}.}
\def\bbone{D. Boer and J. Boomsma, {\PRD{78}{054027}{2008}}.}
\def\bbtwo{D. Boer and J. Boomsma, {\PRD{80}{034019}{2009}}.}
\def\sakai{Y. Sakai, H. Kouno, T. Sasaki and M. Yahiro, {\PLB{705}{349}{2011}}.}
\def\kuonop{T. Sasaki, J.Takahashi, Y. Sakai, H. Kouno, and M. Yahiro, {\PRD{85}{056009}{2012}}.}
\def\dimacp{D. Kharzeev, Annals Phys. {\bf 325}, 205 (2010).}
\def\cme{D. Kharzeev, {\PLB{633}{260}{2006}};
 D. Kharzeev, L. McLerran and H. Warringa, {\NPA{803}{227}{2008}};
 K. Fukushima, D. Kharzeev and H. Warringa, {\PRD{78}{074003}{2008}};
 .}
\def\starexp{B. Abelev et al. [STAR Collaboration], {\PRL{103}{251601}{2009}};
 {\PRC{81}{054908}{2010}}.}

\def\andrianovplb{ A. A. Andrianov,V.A. Andrianov, D.Espriu, X. Planells,
arXiv:1201.3485[hep-ph].}
\def\klevansky{S. Klevansky, Rev. Mod. Phys. {\bf 64}, 649 (1992).}
\def\rehberg{P. Rehberg, S. P. Klevansky and J. Huefner, {\PRC{53}{410}{1996}}.}
\def\amspm{A. Mishra and S. P. Misra, {\ZPC{58}{325}{1993}}.}
\def\hmspmnjl{H. Mishra and S. P. Misra, {\PRD{48}{5376}{1993}}.}
\def\tfd{H. Umezawa, H. Matsumoto and M. Tachiki {\it Thermofield dynamics
and condensed states} (North Holland, Amsterdam, 1982);
P. A. Henning, Phys. Rep. {\bf 253}, 235 (1995).}
\def\amph4{A. Mishra and H. Mishra, {\JPG{23}{143}{1997}}.}
\def\hmcsc{A. Mishra and H. Mishra, {\PRD{74}{054024}{2006}}.}
\def\bccsb{B. Chatterjee, A. Mishra, H. Mishra, {\PRD{84}{014016}{2011}}.}
\def\grosspisarski{D. Gross, R. Pisarski and L. Yaffe, Rev. Mod. Phys. {\bf 53}, 43 (1981).}
\def\hmsanjay{A. Singh, S. Puri and H. Mishra, {\NPA{864}{176}{2011}}.}
\def\tHooft{G. 't Hooft, {\PRD{14}{3432}{1976}}.}
\def\cpcosmo{S. Weinberg,{\PRL{40}{223}{1978}}; F. Wilczeck,{\PRL{40}{279}{1978}}; D. Kharzeev and A. Zhitnitsky,{\NPA{797}{67}{2007}}}
\def\dimapisarski{D. Kharzeev, R.D. Pisarski, M.H.G. Tytgat,{\PRL{81}{512}{1998}}; D. Kharzeev and R.D. Pisarski,
{\PRD{61}{111901}{2000}}.}
\def\bhaswarcp{Bhaswar Chatterjee, Hiranmaya Mishra and Amruta Mishra,{\PRD{85}{114008}{2012}}.}
\def\bhaswarnjlb{Bhaswar Chatterjee, Hiranmaya Mishra and Amruta Mishra,{\PRD{84}{014016}{2011}}.}
\def\chiralB{I.A. Shuspanov and A. V. Smilga, {\PLB{402}{351}{1997}};
N.O. Agasian and I. A. Sushpanov,{\PLB{472}{143}{2000}}; T.D. Cohen,
D.A. McGady, E.S. Werbos,{\PRC{76}{055201}{2007}}; jens O Andersen,
JHEP{\bf 1210},005, (2012).}
\def\providencia{D.P. Menezes, M. Benghi Pinto, S.S. Avancini and C. Providencia
,{\PRC{80}{065805}{2009}}; D.P. Menezes, M. Benghi Pinto, S.S. Avancini , A.P. Martinez
and C. Providencia, {\PRC{79}{035807}{2009}}}
\def\boomsma{J. K. Boomsma and D. Boer, {\PRD{81}{074005}{2010}}}

\def\NJLB{D. Ebert, K.G. Klemenko, M.A. Vdovichenko, A.S. Vshivisev,{\PRD{61}{025005}{2000}}.} 
\def\manfer{E.J. Ferrer, V. de la Incera and C. Manuel, Nucl. Phys.
 B{\bf 747},88 (2006).}
\def\andreasrebhan{F. Preis, A. Rebhan and A. Schmitt, JHEP 1103(2011),033.}
\def\dunc{ R. C. Duncan and C. Thompson, Astrophys. J. 392, L9 (1992).}
\def\duncc {C. Thompson and R. C. Duncan, Astrophys. J. 408, 194 (1993).}
 \def\dunccc{C. Thompson and R. C. Duncan, Mon. Not. R. Astron.  Soc. 275, 255 (1995).}
 \def\duncccc{C. Thompson and R. C. Duncan, Astrophys. J. 473, 322 (1996).}
 \def\kouvel{C. Kouveliotou et al., Astrophys. J. 510, L115 (1999).}
 \def\lat{C. Y. Cardall, M. Prakash, and J. M. Lattimer, Astrophys.  J. 554, 322 (2001).}
 \def\broder{A. E. Broderick, M. Prakash, and J. M. Lattimer, {\PLB{531}{167}{2002}}.}
 \def\lai{D. Lai and S. L. Shapiro, Astrophys. J. 383, 745 (1991).}
\def\fukushimaplb{K. Fukushima, M. Ruggieri and R. Gatto, {\PRD{81}{114031}{2010}}.}
\def\igormag{E.V. Gorbar, V.A. Miransky and I. Shovkovy,{\PRC{80}{032801(R)}{2009}};
ibid, arXiv:1009.1656[hep-ph].}
\def\miranski{V.P. Gusynin, V. Miranski and I. Shovkovy,{\PRL{73}{3499}{1994}};
{\PLB{349}{477}{1995}}; {\NPB{462}{249}{1996}}, E.J. Ferrer and V de la
Incerra,{\PRL{102}{050402}{2009}}; {\NPB{824}{217}{2010}.}}
\def\mukhlb{M. D'Elia, S. Mukherjee and F. Sanfilippo,{\PRD{82}{051501}(2010)}}
\def\florian{Florian Preis, Anton Rebhan and Andreas Schmitt, Lect. Notes Phys. 871 (2013)51-86, arXiv:1204.5077.
} \def\dune{G.Baser, G. Dunne and D. Kharzeev, {\PRL{104}{232301}{2010}}.}
\def\chernodub{M.N. Chernodub, {\PRD{82}{085011}{2010}}.}
\def\frolov{I.E. Frolov, V. Ch. Zhukovsky and K.G. Klimenko, {\PRD{82}{076002}{2010}}.}
\def\nickel{D. Nickel,{\PRD{80}{074025}{2009}}.}
\def\gato{R. Gatto and M. Ruggieri, {\PRD{82}{054027}{2010}}.}
\def\iran{Sh. Fayazbakhsh and N. Sadhooghi, {\PRD{82}{045010}{2010}}.}
\def\hongmag{Deog Ki Hong, arXiv:1010.3923[hep-th].}
\def\digal{T. Mandal, P. Jaikumar and S. Digal, arXiv:0912.1413 [nucl-th] .}
\def\ferrerscmag{E.J. Ferrer, V. de la Incera and C. Manuel, {\PRL{95}{152002}{2005}};
E.J. Ferrer and V. de la Incera, {\PRL{97}{122301}{2006}}; E.J. Ferrer and
V. de la Incera, {\PRD{76}{114012}{2007}}.}
\def\hurwitz{E. Elizalde, J. Phys. {\bf A}:Math. Gen. 18,1637 (1985).}
\def\schaferlat{G.Bali, F. Bruckmann, G. Endrodi, Z. fodor, S.D. Katz etal, JHEP 1202, 044 (2012),1111.4956;
G.S. Bali, F. Bruckmann, G. Endrodi, S.D. Katz and A. Schafer, arXiv:1406.0269 .}
\def\ferreira{M. Ferreira, P. Costa, O. Lorenco, T. Fredrico, C. Providencia, {\PRD{89}{116011}[2014}}.
\def\noronha{E.S. Fraga, J. Noronha and L.F. Palhares, {\PRD{87}{114014}{2013}}.}
\def\fukuprl{K. Fukushima, Y. Hidaka, {\PRL{110}{031601}{2013}}.}
\def\endrodi{F. Bruckmann, G. Endrodi and T. G. Kovacs, JHEP, {\bf 1304}, 112 (2013).}

\end{document}